\documentclass[preprint,11pt,3p]{elsarticle}        
\makeatletter
 \def\ps@pprintTitle{%
 \let\@oddhead\@empty
 \let\@evenhead\@empty
 \def\@oddfoot{}%
 \let\@evenfoot\@oddfoot}
\makeatother

\usepackage[colorlinks,bookmarksopen,bookmarksnumbered,citecolor=red,urlcolor=red]{hyperref}
\usepackage{algorithm,algorithmicx,algpseudocode}
\usepackage{graphicx}
\usepackage{float}
\usepackage{upgreek}
\usepackage{amsmath,amssymb}
\usepackage{subcaption}
\usepackage{rotating}
\usepackage{appendix}
\usepackage{tabularx}
\usepackage{xcolor}
\usepackage{scalerel,stackengine}

\usepackage{xcolor}
\usepackage{xspace}

\usepackage[english]{babel}
\usepackage[utf8]{inputenc}
\usepackage{ulem}  

\DeclareMathAlphabet      {\mathup}{OT1}{\familydefault}{m}{n}

\DeclareMathAlphabet{\mathup}{OT1}{\familydefault}{m}{n}

\DeclareSymbolFont{yhlargesymbols}{OMX}{yhex}{m}{n}
\DeclareMathAccent{\wideparen}{\mathord}{yhlargesymbols}{"F3}
\xdefinecolor{samsi_green}{rgb}{0.04, 0.85, 0.32}
\xdefinecolor{samsi_darkgreen}{rgb}{0.24, 0.7, 0.44}
\xdefinecolor{samsi_mint}{rgb}{0.24, 0.71, 0.54}
\xdefinecolor{samsi_officegreen}{rgb}{0.0, 0.5, 0.0}
\xdefinecolor{samsi_napiergreen}{rgb}{0.16, 0.5, 0.0}
\xdefinecolor{samsi_maroon}{rgb}{0.65,0.06,0.17}
\xdefinecolor{samsi_blue}{rgb}{0,0.2,0.6}
\xdefinecolor{samsi_phthalogreen}{rgb}{0.07,0.21,0.14}
\definecolor{grassgreen}{RGB}{92,135,39}

\newcommand{\floor}[1]{\lfloor #1 \rfloor}
\newcommand{\diag}[1]{\mathsf{diag}\left(#1\right)}                 
\newcommand{\del}[2]{\frac{\partial{#1}}{\partial{#2}}}             
\newcommand{\mat}[1]{\mathbf{{#1}}}                                 
\renewcommand{\vec}[1]{\mathbf{{#1}}}                                 

\newcommand*{\tran}{^{\mkern-1.5mu\mathsf{T}}}                
\newcommand{\trace}{\mathrm{Tr}}                              
\newcommand{\Trace}[1]{\trace \left(#1\right)}                
\newcommand{\norm}[1]{\left\| {#1} \right\|}                  

\newcommand\restr[2]{{ \left.\kern-\nulldelimiterspace        
                     {#1}\vphantom{\big|} \right|_{#2}}}

\newcommand{\Rnum}{\mathbb{R}}  
\newcommand{\x}{\vec{x}}                           
\newcommand{\xa}{{\x}^{\rm a}}                     
\newcommand{\xb}{{\x}^{\rm b}}                     
\newcommand{\xbarb}{\overline{\x}^{\rm b}}         
\newcommand{\xbara}{\overline{\x}^{\rm a}}         

\newcommand{\y}{\mathbf{y}}                        

\newcommand{\Nstate}{\textsc{N}_{\rm state}}                    

\newcommand{\Nobs}{\textsc{N}_{\rm obs}}                        
\newcommand{\Nens}{\textsc{N}_{\rm ens}}                        
\newcommand{\Nsens}{\textsc{N}_{\rm s}}                         
\newcommand{\GM}[2]{\mathcal{N}\!\left( {#1}, {#2}\right)}       

\newcommand{\yk}{\mathbf{y}_k}

\newcommand{\xtin}{\mathbf{x}_0}
\newcommand{\xk}{\mathbf{x}_k}

\stackMath
\newcommand\reallywidehat[1]{%
\savestack{\tmpbox}{\stretchto{%
  \scaleto{%
    \scalerel*[\widthof{\ensuremath{#1}}]{\kern-.6pt\bigwedge\kern-.6pt}%
    {\rule[-\textheight/2]{1ex}{\textheight}}
  }{\textheight}%
}{0.5ex}}%
\stackon[1pt]{#1}{\tmpbox}%
}
\newcommand{\decorrmat}{\mat{C}}                    
\newcommand{\decorrcoeff}{\rho}                     
\newcommand{\locvec}{\vec{L}}                     
\newcommand{\locrad}{l}                     
\newcommand{\inflmat}{\mat{D}}                      
\newcommand{\inflvec}{\boldsymbol\lambda}           
\newcommand{\inflfac}{\lambda}                      

\newcommand{\dates}{{DATeS}\xspace}

\newcommand{\commentout}[1]{\iffalse {#1} \fi}

\begin{document}
%
%
\begin{frontmatter}

	\title{An Optimal Experimental Design Framework for Adaptive Inflation and Covariance Localization for Ensemble Filters}
  
  \author[labela]{Ahmed Attia}
  \author[labela]{Emil Constantinescu} 
	\address[labela]{Mathematics and Computer Science Division,
                   Argonne National Laboratory, Lemont, IL\\
                   Email: attia@mcs.anl.gov
                   }

	\begin{abstract}
    We develop an optimal experimental design framework for adapting the
covariance inflation and localization in data assimilation problems. Covariance inflation and localization are ubiquitously employed to alleviate the effect of using ensembles of finite sizes in all practical data assimilation systems.
The choice of both the inflation factor and the localization radius
can have a significant impact on the performance of the assimilation scheme.
These parameters are generally tuned by trial and error, rendering them expensive to optimize in practice.
Spatially and temporally varying inflation parameter and localization radii have been recently proposed 
and have been empirically proven to enhance the performance of the employed assimilation filter.
In this study, we present a variational framework for adaptive tuning of the inflation and localization parameters.
Each of these parameters is optimized independently, with an objective to minimize the uncertainty in the posterior state.
The proposed framework does not assume uncorrelated observations or prior errors and can in principle be applied without expert knowledge about the model and the observations. Thus, it is adequate for handling dense as well as sparse observational networks.
We present the mathematical formulation, algorithmic description of the approach, and numerical experiments using the two-layer Lorenz-96 model.

  \end{abstract}

	\begin{keyword}
		Sequential data assimilation,
    Ensemble Kalman filter,
    Covariance inflation,
    Covariance localization,
    Optimal experimental design
	\end{keyword}

\end{frontmatter}
%

%
\section{Introduction}
\label{sec:introduction}
  %
  Data assimilation (DA) is a process that combines information from prior knowledge, numerical simulations, and observations, in order to obtain a statistically best estimate of
  the state of a physical system, while accounting for respective errors~\cite{Kalnay_2002_book}.
  The \textit{background} characterizes the best estimate of the true state \textit{prior} to any measurement being available. 
  In applications that involve time-dependent dynamical systems, the
  background at time instant $t_k$ is typically obtained by forward propagation of an initial condition $\x_0$ over the time interval $[t_0,\, t_k]$, by 
  using the model dynamics:
  \begin{equation}\label{eqn:forward_model}
    \xk = \mathcal{M}_{t_0 \rightarrow t_k} (\xtin) \,, \quad k = 1, 2, \ldots \,,
  \end{equation}
  where the model $\mathcal{M}$ typically represents a numerical approximation of differential equations that govern the evolution of the physical system.
  Once an observation is made available, the objective of a DA algorithm is to update the background knowledge and hence produce an analysis that best approximates the unknown truth.
  In the Bayesian context, the analysis is given by the posterior
  distribution of the model state conditioned by the available
  observations. 
  The background errors are usually assumed to follow a Gaussian distribution ${\xb_k - \x_k^{\rm true}}  \in \GM{\vec{0}}{\mat{P}^b_k}$, where $\xb_k\in\Rnum^{\Nstate}$ is the prior (e.g., forecast) state
  and $\mat{P}^b_k\in \Rnum^{\Nstate\times\Nstate}$ is the background error covariance matrix. Here $\x_k^{\rm true}$ is the numerical model representation of the true physical system state at a time instant $t_k$.

  Observations $\y_k \in \mathbb{R}^{\Nobs}$ of the true state are collected at discrete time instants $t_k$,
  \begin{equation}
    \yk \equiv \y(t_k) = \mathcal{H}_k(\x^{\rm true}(t_k)) + \varepsilon_k, \quad k = 0,1,\ldots \,,
  \end{equation}
  where the observation operator $\mathcal{H}_k$ maps a model state $\x_k$ into the observation space.
  The observations are contaminated by measurement and representativeness errors~\cite{Cohn_1997}, which are also assumed to be normally distributed
  $\varepsilon_k \sim \GM{0}{\mat{R}_k}$, where $\mat{R}_k\in \Rnum^{\Nobs\times\Nobs}$ is the observation error covariance matrix at time instant $t_k$.
  Although non-Gaussianity is being actively investigated
  (see, e.g.,~\cite{attia2018ClHMCAtmos,attia2015hmcsmoother,attia2015hmcfilter,attia2016reducedhmcsmoother,hoteit2012particle,van2009particle,stock_nonGaussianDA}),
  assuming Gaussian errors is generally acceptable, and we follow this  assumption  in this work.
  Assuming a linear observation operator and given the Gaussian prior and observation errors, the analysis is also Gaussian $\GM{\xa_k}{\mat{P}^a_k}$, where the analysis mean $\xa_k$ and the analysis covariance matrix $\mat{P}^a_k$ are optimally obtained by applying the
  Kalman filter (KF) update equations~\cite{REKalman_1960a,Kalman_1961}.
  However, exact application of the KF equations in large-scale settings, such as numerical weather prediction and oceanography, is intractable.
  The ensemble Kalman filter (EnKF)~\cite{Evensen_1994} uses a finite ensemble of model states to approximate the exact moments employed in the KF equations, making it attractive for solving large-scale filtering problems.
  The main issues with the EnKF are explained by the introduction of sampling errors, variance underestimation, rank deficiency, and the development of long-range spurious correlations.
  These problems are tackled in practice by applying covariance inflation~\cite{anderson1999monte} and covariance localization~\cite{Houtekamer_2001}. 
  Covariance inflation treats the problem of variance underestimation by inflating the ensemble members around their empirical mean.
  Covariance localization, on the other hand, damps out or completely eliminates spurious correlations between different entries of the model state as the distance between these entries increases.
  This increases the rank of the decorrelated covariance matrix and results in recovery from sampling errors and rank deficiency~\cite{Hamill_2001}.

  The performance of the EnKF implementation depends critically  on the choice of the inflation and localization parameters and therefore must be optimally adjusted.
  Adaptive tuning of the inflation parameter has been proposed in several studies, including~\cite{anderson2007adaptive,Anderson_2009_adaptive_covariance,el2018enhanced,li2009simultaneous}.
  Adaptive covariance localization schemes have also been explored; see, for example,~\cite{anderson2007exploring,bishop2009ensemble_a,bishop2009ensemble_b,flowerdew2015towards,moosavi2018machine}.
  These approaches have been shown to enhance the performance of the EnKF; however, they require selecting further tuning parameters, are computationally expensive, 
  or require expert knowledge about the dynamical model or the observational system~\cite{kirchgessner2014choice}.
  In practice, therefore,  tuning these parameters remains a challenging task carried out mostly by trial and error. Moreover, given the ad hoc nature of such procedures, the mathematical theory is lacking.

  In this work, we introduce a new framework for adaptive tuning of covariance inflation and localization parameters while allowing them to vary over space and in time.
  We follow a gradient-based variational approach, inspired by developments in optimal experimental design (OED) for sensor placement,
  to adaptively tune the space-time-dependent inflation factor and localization radii of influence for covariance localization.
  The proposed algorithms are independent of the specific flavor of the EnKF used for assimilation and can be easily extended to the case of smoothing.
  Moreover, the approach proposed here  inherently accounts for background and observation correlations and hence is suitable for dense as well as sparse  observational networks.
  We use the two-layer Lorenz-96~\cite{lorenz1996predictability} model to test the proposed algorithms, and we compare their performance with an empirically tuned implementation of the EnKF and with the truth.

  The rest of this paper is structured as follows.
  Section~\ref{sec:sequentail_DA} presents a brief overview of sequential data assimilation and the ensemble Kalman filter.
  Section~\ref{sec:OED_inflation_localization} introduces the new framework for adaptive inflation and covariance localization.
  Section~\ref{sec:numerical_experiments} describes the setup of the computational experiments and discusses the numerical results.
  Concluding remarks are presented in Section~\ref{sec:conclusions}

\section{Sequential Data Assimilation}
\label{sec:sequentail_DA}
  Sequential ensemble-based DA filtering algorithms work by repeating two steps: \textit{forecast} and \textit{analysis}.
  In the forecast step, an ensemble of model states is propagated forward by the model equations \eqref{eqn:forward_model} to the next time point where observations are available.
  The result of the forecast step is a forecast ensemble encapsulating the prior knowledge about the model state.
  In the analysis step, the forecast states are corrected, given the collected measurements, to best approximate the truth with lower uncertainty.

  \subsection{Ensemble Kalman Filtering}
  \label{subsec:EnKF}
    The EnKF (see, e.g.,~\cite{Burgers_1998_EnKF,Evensen_1994,Evensen_2003,Houtekamer_1998a}) follows a Monte Carlo approach to approximate the moments of the underlying probability distributions in the Kalman filter equations~\cite{REKalman_1960a,Kalman_1961}.
    At time instant $t_{k-1}$, an ensemble of $\Nens$ states $\{ \xa_{k-1}(e) \}_{e=1,\ldots,\Nens} $ is used to represent the analysis, that is, the posterior probability distribution.
    The model~\eqref{eqn:forward_model} propagates each ensemble state over the time interval $[t_{k-1},t_k]$ to generate a forecast ensemble at $t_k$:
    \begin{subequations}\label{eqn:EnKF_equations}
    \begin{equation}
      \xb_k(e) = \mathcal{M}_{t_{k-1}\rightarrow t_k}(\xa_{k-1}(e)) + \vec{\eta}_k(e),\ \ e=1,\ldots, \Nens . \label{eq:EnKF_forecast}
    \end{equation}

      The model imperfection can be accounted for by adding model errors
      through perturbations to the forecast states. The model errors are generally assumed to be Gaussian random variables, $\vec{\eta}_k \in \GM{\vec{0}}{\mat{Q}_k}$.
    In this study we do not prescribe a $\mat{Q}_k$; however, we account for model errors via inflation, and
    without loss of generality we set $\mat{Q}_k = \mat{0}\,,~ \forall k$.

    The prior, namely, the background state, and the prior covariance matrix are approximated by the mean of the forecast ensemble $\xbarb_k$ and the flow-dependent covariance matrix $\mat{B}_k$,
    respectively, at the next assimilation time instant $t_k$:
    \begin{equation}\label{eqn:EnKF_analysis}
      \begin{aligned}
      \xbarb_k   &=  \frac{1}{\Nens} \sum_{e=1}^{\Nens}{\xb_k(e) } \,, \\
      \mat{X}^{\rm b}_k   &=  [\xb_k(1)- \xbarb_k, \ldots,  \xb_k(\Nens)- \xbarb_k] \,,  \\
      \mat{B}_k   &=  \frac{1}{{\Nens-1}}  \mat{X}^{\rm b}_k \left( \mat{X}^{\rm b}_k \right)\tran  \,. 
      \end{aligned}
    \end{equation}
    \end{subequations}

    Each member of the forecast ensemble $\{ \xb_k(e) \}_{e=1,\ldots,\Nens}$ is analyzed separately by using the Kalman filter formulae~\cite{Burgers_1998_EnKF,Evensen_1994}:
    \begin{equation}
      \label{eqn:EnKF_Analysis_and_gain}
      \begin{aligned}
        \xa_k(e)   &=  \xb_k(e) + \mat{K}_k \left( \left[\yk + \vec{\zeta}_k(e)\right] - \mathcal{H}_k(\xb_k(e)) \right),\ \\
        \mat{K}_k  &=  \mat{B}_k \mat{H}\tran_k { \left(\mat{H}_k \mat{B}_k \mat{H}\tran_k + \mat{R}_k \right)}^{-1}.
      \end{aligned}
    \end{equation}

    The Kalman gain matrix $\mat{K}_k$ uses the linear (or a linearized, e.g., tangent linear) observation operator $\mat{H}_k = \mathcal{H}_k'$, and the same Kalman gain is used for all ensemble members.
    The mean of the posterior ensemble $\xbara_k =  \frac{1}{\Nens} \sum_{e=1}^{\Nens}{\xa_k(e) }$ approximates the true posterior mean and is used as the initial condition for future predictions.
    The posterior covariance matrix $\mat{A}_k$ approximates the true posterior covariance matrix $\mat{P}_k^a$ and is given by
    \begin{eqnarray}\label{eqn:postior_covariance}
      \mat{A}_k = \left( \mat{I} - \mat{K}_k \mat{H} \right) \mat{B}_k
      \equiv \left( \mat{B}_k^{-1} + \mat{H}_k\tran \mat{R}^{-1} \mat{H}_k \right)^{-1}.
    \end{eqnarray}

    The stochastic ``perturbed observations'' version~\cite{Burgers_1998_EnKF} of the ensemble Kalman filter adds a different realization
    of the observation noise $\vec{\zeta}_k \sim \GM{\vec{0}}{\mat{R}_k}$ to each individual assimilation.
    On the other hand, square root ``deterministic'' formulations of the EnKF~\cite{Tippett_2003_EnSRF} avoid adding random noise to observations and thus avoid additional sampling errors.
    For additional discussion of the EnKF variants, see, for example,~\cite{asch2016data,Evensen_2007_book}.
    We note that the discussion hereafter is independent of the specific flavor of the EnKF used in the analysis step of the filter.
    In what follows, we drop the time subscript $k$, for brevity of notation, and assume that all calculations are carried out at time instant $t_k$  unless otherwise stated explicitly.
    %

  \subsection{Sampling errors, spurious correlations, and filter divergence}
  \label{subsec:EnKF_limitations}
    Spurious correlations over long spatial distances occur when using
    a small ensemble of model states to approximate the covariances~\cite{Evensen_2007_book}.
    Spurious correlations, model errors, and forcing the Gaussian framework on non-Gaussian settings are known to lead to underestimation of the true state variances~\cite{Furrer_2007_covEstimation}
    and consequently can lead to filter divergence.
   
    In practice, the problem of covariance underestimation in the EnKF is
    alleviated by applying some form of covariance inflation. For example, one can scale the forecast (or the analysis) ensemble around its mean~\cite{anderson1999monte}.
    On the other hand, spurious correlations are reduced by applying covariance localization~\cite{Houtekamer_2001}. In geoscientific DA applications, distance-based covariance localization is widely used.
    In order to apply covariance inflation, an \textit{inflation factor} must be provided. Also, in order to apply distance-based covariance localization, a radius of influence (e.g., \textit{localization radius}) is required. 
    Both the inflation factor and the localization radius could be
    fixed scalars, in other words, they can be made space and time independent.
    On the other hand, making each of these parameters space and/or time dependent can enhance the performance of the filter~\cite{anderson2007adaptive,Anderson_2009_adaptive_covariance,el2018enhanced,flowerdew2015towards,Herschel_1999a}.
    Adaptive tuning of the inflation factor~\cite{anderson2007adaptive,anderson1999monte} and localization radius~\cite{anderson2007adaptive,Hamill_2001,Houtekamer_2001,Houtekamer_2005} is
    a nontrivial problem, especially in space-time--dependent settings~\cite{Constantinescu_A2007b,Constantinescu_A2007c}.
    Inflation and covariance localization are discussed in detail in~\S\ref{subsec:inflation}, \S\ref{subsec:localization}.

  \subsection{Inflation}
    \label{subsec:inflation}
    As discussed in~\S\ref{subsec:EnKF_limitations}, the problem of filter divergence due to underestimated ensemble covariance is mitigated by applying covariance inflation.
    The effect of this procedure is that the ensemble background covariance matrix $\mat{B}$ is replaced with an inflated version $\widetilde{\mat{B}}$.
    The most popular forms of covariance inflation can be classified into two types: \textit{additive} and \textit{multiplicative} inflation.
    In additive covariance inflation, a diagonal matrix $\inflmat = \diag{\inflvec}$, where $\inflvec=\left(\inflfac_1, \inflfac_2,\ldots, \inflfac_{\Nstate} \right)\tran$, is added to the ensemble covariance matrix;
    that is, $\widetilde{\mat{B}} = \inflmat + \mat{B}$. Here, the elements on the diagonal of $\inflmat$ are slightly larger than zero, namely, $0 \leq \inflfac_i \leq \inflfac^{\rm u}$ for some upper limit $\inflfac^{\rm u}$.
    The inflation factor $\inflfac_i$ can be held constant for all grid points or can be varied, in other words, made space dependent~\cite{Anderson_2009_adaptive_covariance,Herschel_1999a}.
    %
    
      Multiplicative inflation, on the other hand, works by pushing the ensemble members away from the ensemble mean by a given inflation factor.
      Assume that the inflation factors are held constant over the model space domain, namely, $\inflfac_i=\inflfac\,,~\forall\, i=1,2,\ldots, \Nstate$.
      The inflated covariance matrix $\widetilde{\mat{B}}$ is obtained by magnifying the ensemble of forecast anomalies as follows,
      \begin{equation}\label{eqn:multiplicative_inflation}
        \begin{aligned}
          \widetilde{\mat{X}^{\rm b}} &= \left[\sqrt{\inflfac}\left(\xb(1)- \xbarb\right), \ldots,  
            \sqrt{\inflfac}\left(\xb(\Nens)- \xbarb\right) \right] \,,  \\
          \widetilde{\mat{B}}   &=   \frac{1}{\Nens-1}  \widetilde{\mat{X}^{\rm b}} 
            \left( \widetilde{\mat{X}^{\rm b}} \right)\tran  = \inflfac \, \mat{B}  \,,
        \end{aligned}
      \end{equation}
      where $1\leq \inflfac^l \leq \inflfac \leq \inflfac^u$ for some lower $\inflfac^l$ and upper  $\inflfac^u$ bounds, respectively.
      This formulation, however, restricts the inflation factor to be
      identical for all state vector entries. Arguably, the inflation should be made more flexible by allowing spatial variability~\cite{anderson2007adaptive}.
      This generalization is formulated as follows. Consider the inflation matrix $\inflmat :=\diag{\inflvec}= \sum_{i=1}^{\Nstate}{\inflfac_i \vec{e}_i \vec{e}_i\tran}$, with
      $ \inflvec = \left( \inflfac_1, \inflfac_2, \ldots, \inflfac_{\Nstate} \right) \tran $, where $\inflfac^l_i \leq \inflfac_i \leq \inflfac^u_i$ for some lower and upper bounds on the inflation factor 
      for each entry of the state vector.
      The matrix of the magnified ensemble anomalies and the inflated ensemble-based covariance matrix $\widetilde{\mat{B}}$ respectively are the following:
      \begin{equation}\label{eqn:space_multiplicative_inflation}
        \begin{aligned}
          \widetilde{\mat{X}^{\rm b}} &= \mat{D}^{\frac{1}{2}} \mat{X}^{\rm b} \,,  \\
          \widetilde{\mat{B}} &= \frac{1}{\Nens-1}  \widetilde{\mat{X}^{\rm b}} \left( \widetilde{\mat{X}^{\rm b}} \right)\tran 
          = \mat{D}^{\frac{1}{2}}  \left( \frac{1}{\Nens-1} \mat{X}^{\rm b} \left( \mat{X}^{\rm b} \right)\tran \right) \mat{D}^{\frac{1}{2}}
          = \mat{D}^{\frac{1}{2}}  \mat{B} \mat{D}^{\frac{1}{2}} \,.
        \end{aligned}
      \end{equation}

      Whether additive or multiplicative inflation is applied, the
      inflated covariance matrix is used in the analysis step of the
      filter, and thus it changes the posterior ensemble covariance matrix.
      The inflated Kalman gain $\widetilde{\mat{K}}$ and inflated analysis error covariance matrix $\widetilde{\mat{A}}$, respectively, can be written as
      \begin{subequations} \label{eqn:infl_post_gain_cov}
      \begin{align} 
        \widetilde{\mat{K}} &= \widetilde{\mat{B}} \mat{H}\tran { \left(\mat{H} \widetilde{\mat{B}} \mat{H}\tran + \mat{R} \right)}^{-1} \label{eqn:infl_post_gain}\,,  \\ 
        \widetilde{\mat{A}} &= \left( \mat{I} - \widetilde{\mat{K}} \mat{H} \right) \widetilde{\mat{B}}
        \equiv \left( \widetilde{\mat{B}}^{-1} + \mat{H}\tran \mat{R}^{-1} \mat{H} \right)^{-1} \,. \label{eqn:infl_post_cov}  
      \end{align}
      \end{subequations}

      Without loss of generality, hereafter we assume multiplicative inflation is used.
      In order to apply covariance inflation, that is,  Equation~\eqref{eqn:multiplicative_inflation} or~\eqref{eqn:space_multiplicative_inflation},
      the inflation factor $\inflvec$ must be tuned for the
      application in hand in order to prevent the ensemble collapse. 
      In \S\ref{sec:OED_inflation_localization}, we propose an approach to adaptively tune the space-time covariance inflation parameter $\inflvec$.

  \subsection{Covariance localization}
  \label{subsec:localization}
    In order to alleviate spurious correlations developed due to small ensemble size, covariance localization ~\cite{Hamill_2001,Houtekamer_2001,Whitaker_2002a} is performed by applying a pointwise multiplication to the ensemble
    covariances~\cite{bernstein2005matrix,Horn_1990_Hadamard,million2007hadamard,schur1911bemerkungen} with a decorrelation or localization  matrix $\decorrmat$.
    In its standard form~\cite{bishop2007flow,Hamill_2001}, covariance localization is carried out in the model state space. 
    Specifically, the ensemble covariance matrix $\mat{B}$ is replaced in the analysis step of the filter with a localized version $\widehat{\mat{B}} = \mat{B}\odot \decorrmat$, 
    where $\decorrmat \in \Rnum^{\Nstate \times \Nstate}$ is a spatial decorrelation matrix and $\odot$ refers to the Hadamard (Shur) pointwise product of matrices~\cite{bernstein2005matrix,Horn_1990_Hadamard,million2007hadamard,schur1911bemerkungen}.
    The entries of the decorrelation matrix $\decorrmat$ can be calculated, for example, by using  Gauss-like formulae, such as a Gaussian kernel~\eqref{eqn:Gauss_decorr},
    or the fifth-order piecewise-rational function~\eqref{eqn:Gaspari_Cohn} of Gaspari and Cohn~\cite{Gaspari_1999_correlation}, hereafter called the GC function:
    \begin{subequations}\label{eqn:localization_functions}
      \begin{equation}\label{eqn:Gauss_decorr}
        \decorrcoeff_{i,j} (L)  = \exp{\left( \frac{-d(i,j)^2}{2 L^2 } \right)} \,, 
      \end{equation}
      \begin{equation}\label{eqn:Gaspari_Cohn}
      \resizebox{0.90\hsize}{!}{$
        \decorrcoeff_{i,j} (L) =
        \begin{cases}
          - \frac{1}{4} \left( \frac{d(i,j)}{L} \right)^5 + \frac{1}{2} \left( \frac{d(i,j)}{L} \right)^4 
            + \frac{5}{8} \left( \frac{d(i,j)}{L} \right)^3 - \frac{5}{3} \left( \frac{d(i,j)}{L} \right)^2 + 1\,, \quad  & 0 \leq d(i,j) \leq L \\
          \frac{1}{12} \left( \frac{d(i,j)}{L} \right)^5 - \frac{1}{2} \left( \frac{d(i,j)}{L} \right)^4 
            +  \frac{5}{8} \left( \frac{d(i,j)}{L} \right)^3 + \frac{5}{3} \left( \frac{d(i,j)}{L} \right)^2 
              - 5 \left( \frac{d(i,j)}{L} \right)+4-\frac{2}{3} \left(\frac{L}{d(i,j)} \right)\,,  \quad & L \leq d(i,j) \leq 2L \\
          0\,, \quad & 2L \leq d(i,j)\,,
        \end{cases}
      $}
      \end{equation}
    \end{subequations}
    where $L$ is the localization length scale (i.e., localization radius) and $d(i, j)$ is the distance between the $i$th and  $j$th grid points.
    Unlike the Gaussian formula~\eqref{eqn:Gauss_decorr}, the GC
    function~\eqref{eqn:Gaspari_Cohn} is designed to have compact support such that it is nonzero only
    for a small local region and zero everywhere else~\cite{petrie2008localization}.
    Specifically, this correlation function attenuates covariance
    entries with increasing distance and has a compact support of $2L$.
    Obviously, the localization coefficient $\decorrcoeff$ is a function of the localization radius $L$.
    The distance metric $d(i,j)$ depends on the problem and is
    application specific. It
    can be  the distance between two state grid points, between two observations,
    or  between a model grid point and an observation.
    
    The correlation length scale $L$ controls the correlation function~\cite{lorenc2003potential} and must be adjusted for the application at hand.
    As suggested in~\cite{kepert2009covariance,Mitchell_2002a}, the radius of influence $L$ should be allowed to be space-time dependent and consequently must be tuned for each entry of the
    state vector.
    For example, the localization coefficients for Gaussian localization can take the form
    \begin{equation}\label{eqn:localization_matrix_Gauss}
      \decorrcoeff_{i,j} (\locrad_{i,j}) = \exp{\left(- \frac{d(i, j)^2}{2 \locrad_{i,j}^2 } \right)} \,,
    \end{equation}
    where $l_{i,j}$ is the localization radius corresponding to the $i$th and $j$th grid points.
    In general, one can parametrize the global space-dependent localization matrix $\mat{C}$ in various ways; however, to keep the presentation simple, 
    we will maintain a high-level description of its entries:
    \begin{equation}\label{eqn:localization_matrix}
      \decorrmat = [\decorrcoeff_{i,j} (\locrad_{i,j})]_{i,j=1,2,\ldots, \Nstate} \,.
    \end{equation}

    By localizing the ensemble covariance matrix, the posterior ensemble covariance matrix is altered.
    From~\eqref{eqn:EnKF_Analysis_and_gain}, the localized Kalman gain $\widehat{\mat{K}}$ and the posterior covariance matrix $\widehat{\mat{A}}$, respectively, become
      \begin{subequations}
      \begin{eqnarray}
        \widehat{\mat{K}} &=& \widehat{\mat{B}} \mat{H}\tran { \left(\mat{H} \widehat{\mat{B}} \mat{H}\tran + \mat{R} \right)}^{-1}  \,, \label{eqn:localized_gain}  \\
          \widehat{\mat{A}} &=& \left( \mat{I} - \widehat{\mat{K}} \mat{H} \right) \widehat{\mat{B}}
      \equiv \left( \widehat{\mat{B}}^{-1} + \mat{H}\tran \mat{R}^{-1} \mat{H} \right)^{-1} \label{eqn:localized_postior_covariance} \,.
      \end{eqnarray}
      \end{subequations}

    This form of localization is commonly referred to as \textit{$\mat{B}$-localization}.
    In practice, however, applying localization directly to the background error covariance matrix $\mat{B} \in \Rnum^{\Nstate \times \Nstate}$
    is computationally prohibitive because of the high dimensionality of the model state space.
    An alternative approach is to localize the effect of assimilated observations to neighboring grid points. 
    
    The latter approach is commonly referred to as~\textit{$\mat{R}$-localization}.
      In this approach the covariance localization is carried out in the observation
      space~\cite{chen2010cross,bishop2007flow,bishop2009ensemble_a,bishop2009ensemble_b}.
      This is typically done by decorrelating $\mathbf{H} \mat{B}$, and possibly $\mathbf{HB}\mathbf{H}\tran$.
      In this version of localization, $\mat{HB} $ is replaced with
      $\widehat{\mat{HB} } = \decorrmat^{\rm loc, 1} \odot \mat{HB} $, and $\mat{HB} \mat{H}\tran$
      is possibly replaced with $\reallywidehat{\mat{HB} \mat{H}\tran} = \decorrmat^{\rm loc, 2} \odot \mat{HB} \mat{H}\tran $.
      The matrices $\decorrmat^{\rm loc, 1} \in \Rnum^{\Nobs \times \Nstate}$ and $\decorrmat^{\rm loc, 2} \in \Rnum^{\Nobs \times \Nobs}$
      are two localization matrices.
      Here $\decorrmat^{\rm loc, 1}$ is constructed based on distances between observations and model grid points.
      On the other hand, $\decorrmat^{\rm loc, 2}$ relies on distances between pairs of observation grid points.
      To apply $\mathbf{R}-$localization, one can localize $\mathbf{H} \mat{B}$ only or apply localization to
      both $\mathbf{H} \mat{B}$ and $\mathbf{H} \mat{B} \mathbf{H}\tran$.
      Based on the chosen approach, the localized posterior covariance matrix $\widehat{\mat{A}}$
      can be written in the following respective forms:
      \begin{equation}
        \widehat{\mat{A}} = \mat{B}-\widehat{\mat{HB} }\tran { \left( \mat{R} + {\mat{HB} \mat{H}\tran} \right)}^{-1} \widehat{\mat{HB} }
        \quad \text{or} \quad
        \widehat{\mat{A}} = \mat{B}-\widehat{\mat{HB} }\tran { \left( \mat{R} + \reallywidehat{\mat{HB} \mat{H}\tran} \right)}^{-1} \widehat{\mat{HB} }
        \,.
      \end{equation}

      The space-dependent localization radii $\locvec$ must be tuned if $\mat{B}$ or $\mat{R}$ localization is carried out.
      In \S\ref{sec:OED_inflation_localization}, we propose an approach for automatically tuning this parameter.

\section{OED Approach for Adaptive Inflation and Covariance Localization}
\label{sec:OED_inflation_localization}
  The main goal of solving a DA problem is to obtain a posterior state that approximates the truth with minimum uncertainty.
  Since KF equations are derived to generate a minimum-variance
  unbiased estimator (MVUE) of the truth and borrowing the idea from PDE-based OED literature (see, e.g.,
  \cite{AlexanderianPetraStadlerEtAl14,AlexanderianPetraStadlerEtAl16,attia2018goal,CrestelAlexanderianStadlerEtAl17,HaberHoreshTenorio08,HaberMagnantLuceroEtAl12,TenorioLuceroBallEtAl13}),
  the inflation and localization parameters should be tuned to yield minimum uncertainty of the posterior state.
  Here, we propose an OED-based approach to tune the space-time inflation and localization parameters to produce an analysis with minimum uncertainty.
  An OED application requires defining a scalar measure of posterior uncertainty, which is used to define the optimality criterion.
  Two popular design criteria are the Bayesian A- and D-optimality~\cite{AtkinsonDonev92,ChalonerVerdinelli95,Pukelsheim93}.
  A-optimality seeks to minimize the trace of the Gaussian posterior covariance matrix, whereas D-optimality minimizes its determinant.
  In this work, we focus on using A-optimality as the main OED criterion for both adaptive inflation~\S\ref{subsec:OED_inflation} and adaptive localization~\S\ref{subsec:OED_localization}.

  %
  \subsection{OED adaptive inflation}
    \label{subsec:OED_inflation}
    The OED adaptive inflation problem is described by the following
    minimization problem,
    \begin{equation} \label{eqn:multiplicative_inflation_A_opt}
      \begin{aligned}
      & \min_{\inflvec \in \Rnum^{\Nstate}}{ \Psi^{\rm Infl} (\inflvec) - \alpha \, \Phi(\inflvec) }  \\
        \text{subject to}\quad & 1 \leq \inflfac^l_i \leq \inflfac_i \leq \inflfac^u_i, \quad i = 1, \ldots, {\Nstate} \,,
      \end{aligned}
    \end{equation}
    where $\Psi^{\rm Infl}$ is the specific design criterion,
    $\Phi(\inflvec): \Rnum_{+}^{\Nsens} \mapsto [0, \infty)$ is a regularization function, and $\alpha > 0$ is a user-defined regularization parameter.
    Here, $\inflfac^l_i$ and $\inflfac^u_i$ are predefined lower and upper bounds on the inflation factors.
    For a discussion on choosing inflation bounds, see for example~\cite{luo2013covariance}.

    \paragraph{The A-optimal criterion}
      We define the A-optimal inflation parameter $\inflvec^{\rm A-OED}$ as the one that minimizes the posterior trace~\eqref{eqn:infl_post_cov}, and hence
      the A-optimality criterion becomes
      \begin{equation}\label{eqn:multiplicative_inflation_A_opt_goal}
        \Psi^{\rm Infl}(\inflvec) := \Trace{\widetilde{\mat{A}}(\inflvec)} \,,
      \end{equation}
      where $\widetilde{\mat{A}}$ is a function of the inflation parameter $\inflvec$ and hence is stated explicitly in~\eqref{eqn:multiplicative_inflation_A_opt_goal}.
      Since the optimality criterion~\eqref{eqn:multiplicative_inflation_A_opt_goal} is a sum of positive values (i.e., posterior variances) and since the inflation state space
      lies in the non-negative side of the real line, we choose the
      $\ell_1$ norm for regularization. 
      Specifically, we set
      \begin{equation} \label{eqn:regularization_function}
        \Phi(\inflvec):=\norm{\inflvec - \vec{1}}_1 = \sum_{i=1}^{\Nstate}{\left| \inflfac_i-1 \right| }  \,,
      \end{equation}
      where $\vec{1} \in \Rnum^{\Nstate}$ is a vector with all entries equal to $1$. This regularization norm amounts to the total inflation applied to each member of the forecast ensemble of anomalies.

      We note that we choose the sign of the regularization term in~\eqref{eqn:multiplicative_inflation_A_opt} to be negative, unlike the traditional formulation of OED problems,
      to force covariance inflation by pushing the inflation factors away from the predefined lower bound.
      To better understand this idea, consider the case where the background errors and observation errors are both uncorrelated, with variances $\sigma_i^2$ and $r_i^2$, respectively.
      Assuming an identity observation operator, namely,
      $\mathcal{H}=\mat{H}=\mat{I}$, then the  optimality criterion in this case reduces to
      \begin{equation}
        \Psi^{\rm Infl}(\inflvec) := \Trace{ \widetilde{\mat{A}} }  = \sum_{i=1}^{\Nstate}{\left(\inflfac_i^{-1} \sigma_i^{-2}  +r_i^{-2} \right)^{-1} } \,.
      \end{equation}

      Therefore, decreasing $\inflfac_i $ reduces the value of the objective $\Psi^{\rm Infl}$, which in turn means that the optimizer will often
      move toward the lower bound on $\inflvec$. The purpose of regularization here is to allow for inflation, while maintaining the posterior trace as small as possible.
      This requires pushing the inflation factors away from the lower bound. Consequently, since the regularization parameter $\alpha$ is assumed to be non-negative, we choose a negative sign for the regularization term.

    \paragraph{Solution of the A-optimality optimization problem}
    A gradient-based approach is typically followed for solving~\eqref{eqn:multiplicative_inflation_A_opt}.
    The gradient of the objective in~\eqref{eqn:multiplicative_inflation_A_opt} is summarized by 
    %
    $  \nabla_{\inflvec}{ \left(\Psi^{\rm Infl} (\inflvec) - \alpha \, \Phi(\inflvec) \right)} 
      = \nabla_{\inflvec}{ \Psi^{\rm Infl} (\inflvec)} - \alpha \, \nabla_{\inflvec}{ \Phi(\inflvec) } \,,
      $ 
    %
    which requires the derivatives of both the optimality criterion and the regularization term.
    Starting with the A-optimality criterion and given that $\widetilde{\mat{B}} = \mat{D}^{\frac{1}{2}}  \mat{B} \mat{D}^{\frac{1}{2}}$,
    one can evaluate $\Psi^{\rm Infl} (\inflvec) = \Trace{ \widetilde{\mat{A}} }$ as follows,
    \begin{equation}\label{eqn:Infl_A_Opt_Objective}
      \begin{aligned}
        \Trace{ \widetilde{\mat{A}} }
        &= \Trace{ \left( \widetilde{\mat{B}}^{-1} + \mat{H}\tran \mat{R}^{-1} \mat{H} \right)^{-1} }
        %
        = \Trace{ \widetilde{ \mat{B}}}
          - \Trace{ \mat{G}^{-1}  \mat{H} \widetilde{\mat{B}}  \widetilde{ \mat{B} } \mat{H}\tran
          } \,,
      \end{aligned}
    \end{equation}
    where $\mat{G} = \mat{R} + \mat{H} \widetilde{\mat{B}} \mat{H}\tran $.
    Here, the Sherman-Morrison-Woodbury
    formula~\cite{sherman1950adjustment,woodbury1950inverting}, the
    cyclic property, and the fact that the matrix trace is a linear
    operator are utilized. 
    The first term in~\eqref{eqn:Infl_A_Opt_Objective} is the total variance of the inflated ensemble, which can be
    easily calculated given the forecast ensemble.
    The second term is the trace of a square matrix of dimension $\Rnum^{\Nobs} \times \Rnum^{\Nobs}$.
    This is typically small in data assimilation applications, compared with the model state space dimension.
    
    The gradient of~\eqref{eqn:Infl_A_Opt_Objective} is derived
    in~\S\ref{append:A_inflation_gradient} and is given by
    \begin{equation}\label{eqn:Infl_A_Opt_Grad_full}
      \nabla_{\inflvec}{\Trace{\widetilde{\mat{A}}}}
      = \sum_{i=1}^{\Nstate}
        {\frac{1}{\inflfac_i} \,\vec{e}_i \, \vec{e}_i\tran\, \left(
          \vec{e}_i \tran \widetilde{\mat{B}} 
            - \mat{H}\tran \mat{G}^{-1}  \mat{H}  \widetilde{\mat{B}}  \widetilde{\mat{B}}
            + \left( \mat{H}\tran \mat{G}^{-1}  \mat{H}  \widetilde{\mat{B}}
              - \mat{I} \right) \widetilde{\mat{B}} \mat{H}\tran \mat{G}^{-1} \mat{H} \widetilde{\mat{B}}
          \right)\, \vec{e}_i}\,,
    \end{equation}
    and the gradient of the regularization term depends on the specific choice of the regularization norm. 
    Since we choose $\ell_1$ norm for regularization as defined by~\eqref{eqn:regularization_function}, it follows that $\nabla_{\inflvec}{\Phi(\inflvec) = \vec{1}}$.

    We conclude this section with a brief summary.
    The objective of the OED adaptive inflation procedure is to find an optimal inflation factor for the current assimilation cycle where the analysis step of the filter is carried out.
    This is done as follows. Once the forecast ensemble is made available, the optimization problem~\eqref{eqn:multiplicative_inflation_A_opt}, 
    is solved by using a gradient-based optimization procedure to find the A-optimal inflation factor $\inflvec^{\rm A-OED}$.
    The inflated background covariance matrix~\eqref{eqn:space_multiplicative_inflation} is calculated by using $\inflvec=\inflvec^{\rm A-OED}$ and is used in the analysis step of the filter.
    The resulting analysis ensemble is then integrated forward in time (i.e., the forecast step) to the next assimilation  cycle. 

  \subsection{OED adaptive covariance localization}
  \label{subsec:OED_localization}
    Following the approach described in~\S\ref{subsec:OED_inflation}, here we describe an OED A-optimality approach to tune the localization radii of the covariance localization.
  
    \subsubsection{State-space formulation}
    \label{subsubsec:OED_localization_state}
      Assume $\locvec = \left( \locrad_1, \locrad_2, \ldots, \locrad_{\Nstate} \right) \tran $ is a vector composed of the localization radii associated with the model grid points 
      and is used to construct the localization matrix $\decorrmat \equiv \decorrmat (\locvec) $. 
      The OED adaptive localization problem is described by the optimization problem
      \begin{equation} \label{eqn:B_localization_opt}
        \begin{aligned}
        & \min_{\locvec \in \Rnum^{\Nstate}}{ \Psi^{\rm B-Loc} (\locvec) +  \gamma \, \Phi(\locvec) }  \\  
          \text{subject to}\quad & 0< \locrad^l_i \leq \locrad_i \leq \locrad^u_i, \quad i = 1, \ldots, {\Nstate} \,,
        \end{aligned}
      \end{equation}
      where $\Psi^{\rm B-Loc}$ is the design criterion and $\locrad^l_i$ and $\locrad^u_i$ are the bounds on the localization radii.
      Again, as in the inflation problem formulation, $\Phi(\locvec): \Rnum_{+}^{\Nstate} \mapsto [0, \infty)$ is a regularization function, and $\gamma > 0$ is a user-defined penalty parameter.
     
      The OED A-optimality criterion for $\mat{B}-$localization is defined to be the trace of the decorrelated posterior covariance matrix~\eqref{eqn:localized_postior_covariance}; 
      that is, $\Psi^{\rm B-Loc}(\locvec) := \Trace{ \widehat{\mat{A}}(\locvec) }$.
      Moreover, following the same argument in the adaptive inflation procedure, we choose the $\ell_1$ norm as a regularization function; that is, we set $\Phi(\locvec):=\norm{\locvec}_1 = \sum_{i=1}^{\Nstate}{\locrad_i} $.
      The A-optimal localization radii vector $\locvec^{\rm A-OED}$ is defined as the solution of the optimization problem
      \begin{equation} \label{eqn:B_localization_A_opt}
        \begin{aligned}
        & \min_{\locvec \in \Rnum^{\Nstate}}{ \Trace{ \widehat{\mat{A}}(\locvec) } +  \gamma \, \sum_{i=1}^{\Nstate}{\locrad_i} }  \\  
          \text{subject to}\quad & 0 < \locrad^l_i \leq \locrad_i \leq \locrad^u_i, \quad i = 1, \ldots, {\Nstate} \,.
        \end{aligned}
      \end{equation}

      Now, we turn our attention to the evaluation of the A-optimality criterion, that is, the first term in~\eqref{eqn:B_localization_A_opt}, and its derivative.
      The A-optimality criterion is
      \begin{equation}\label{eqn:B_localization_objective}
          \Psi^{\rm B-Loc}(\locvec) 
           := \Trace{ \widehat{\mat{A}} }
            = \Trace{ \widehat{\mat{B}} } - \Trace{ \left( \mat{R} + \mat{H} \widehat{\mat{B}} \mat{H}\tran \right)^{-1} \mat{H} \widehat{\mat{B}} 
            \widehat{\mat{B}} \mat{H}\tran  }\,,
      \end{equation}
      where, as before,  $\widehat{\mat{B}} = \mat{B}\odot \decorrmat$ and $\decorrmat \in \Rnum^{\Nstate \times \Nstate}$.
      
      The first term in~\eqref{eqn:B_localization_objective} is the total variance of the localized ensemble, which is equal to $\Trace{ \mat{B} }$, because all diagonal entries of $\decorrmat$ are equal to one.
      The gradient of~\eqref{eqn:B_localization_objective} is clearly defined based on the choice of the decorrelation matrix $\decorrmat$.
      Given the localization functions~\eqref{eqn:localization_functions}, one can formulate a global space-dependent localization matrix $\decorrmat$ in many ways.
      The discussion presented here can be applied to other forms; however, for clarity and other reasons explained below, we focus on the following form:
      \begin{equation}\label{eqn:decorrelation_matrix}
        \decorrmat :=
          \frac{1}{2} \left( \decorrmat_r + \decorrmat_c \right)
          = \frac{1}{2} \left( \decorrmat_r + \decorrmat_r\tran \right)
          = \frac{1}{2} \left[\decorrcoeff_{i,j} (\locrad_i) +  \decorrcoeff_{i,j} (\locrad_j)  \right] _{i,j=1,2,\ldots,\Nstate}\,.
      \end{equation}

      The matrix in the first term, $\decorrmat_r = \left[ \decorrcoeff_{i,j} (\locrad_i)  \right] _{i,j=1,2,\ldots,\Nstate} $,
      fixes the radius of influence for each row of the decorrelation matrix,
      while the matrix $\decorrmat_c = \decorrmat_r\tran$ fixes the radius of influence over the columns of the localization
      matrix.
      The average in~\eqref{eqn:decorrelation_matrix}
      preserves the symmetry of the localized covariance matrix and thus avoids complications in the optimization step.
        However, the decorrelation matrix~\eqref{eqn:decorrelation_matrix} is not necessarily positive definite, 
        and thus the localized covariance matrix is not guaranteed to be positive definite.
        To guarantee that the posterior covariance matrix is positive definite, 
        one needs to consider positive definite localization kernels. 
        Consider, for example, the localization matrix $\overline{\decorrmat_d}={\decorrmat_d}\tran \, {\decorrmat_d}$, where 
        $\decorrmat_d = \left[ \decorrcoeff_{i,j} (\locrad_{\min{(i,j)}})  \right]_{i,j=1,2,\ldots,\Nstate }$.
        A symmetric positive definite localization kernel can then take the form
        $\decorrmat = \diag{\overline{\decorrmat_d}}^{\frac{1}{2}} \, \overline{\decorrmat_d} \, 
          \diag{\overline{\decorrmat_d}}^{\frac{1}{2}}$.
        However, this process may be too expensive to compute in practice.

      Given our specific choice of the localization kernel~\eqref{eqn:decorrelation_matrix} and as derived in~\S\ref{append:A_localization_gradient}, the gradient of $\Trace{ \widehat{\mat{A}} }$ takes the form
      \begin{subequations}
      \label{eqn:B_localization_derivatives}
      \begin{equation} \label{eqn:B_localization_derivative}
        \nabla_{\locvec}{\Psi^{\rm B-Loc} } 
          = \nabla_{\locvec}{ \Trace{ \widehat{\mat{A}} } } 
          = \sum_{i=1}^{\Nstate}{ 
            \vec{e_i} 
            \,  
            \mat{l}_{B,i}  \left( \mat{I} + \mat{H}\tran \mat{R}^{-1} \mat{H} \widehat{\mat{B}} \right)^{-1}
            \,
            \left( \mat{I} + \widehat{\mat{B}}  \mat{H}\tran \mat{R}^{-1} \mat{H} \right)^{-1} \vec{e}_i  
            }  \,,  \\
      \end{equation}
      with $\vec{e}_i$ being the $i$th cardinality vector in $\Rnum^{\Nstate}$,
      and
      \begin{align}\label{eqn:B_loc_strides}
        \mat{l}_{B,i} &= \mat{l}_i\tran \odot  \left( \vec{e}_i\tran \mat{B} \right)  \\
        \mat{l}_i 
          &= \left( \del{\decorrcoeff_{i,1}(\locrad_i)}{\locrad_i},
                    \del{\decorrcoeff_{i,2}(\locrad_i)}{\locrad_i},
                    \ldots,  
                    \del{\decorrcoeff_{i,\Nstate}(\locrad_i)}{\locrad_i} 
             \right) \tran \,,
      \end{align}
      \end{subequations}
      where $\del{\decorrcoeff_{i,j}(\locrad_i)}{\locrad_i}$ is the derivative of the chosen localization function.
      For example, the derivatives of both Gauss and GC localization functions are given by~\eqref{eqn:loc_fun_derivative_Gauss}, and~\eqref{eqn:loc_fun_derivative_GC}, respectively,
      and the derivative of the regularization term is $\nabla_{\locvec}(\gamma \, \norm{\locvec}_1) = \gamma \vec{1}$. This result, together with 
      the derivative described by~\eqref{eqn:B_localization_derivatives}, formulates the derivative of the objective function in~\eqref{eqn:B_localization_A_opt}.
      
      In summary, the objective of the OED adaptive localization  is to find an optimal space-dependent vector of localization radii for the current assimilation cycle of the filtering procedure.
      Specifically, given the forecast ensemble, the optimization
      problem~\eqref{eqn:B_localization_A_opt} is solved, resulting in
      $\locvec^{\rm A-OED}$. 
      The decorrelated covariance matrix $\widehat{\mat{B}}=\decorrmat(\locvec) \odot \mat{B}$ is calculated by using $\locvec=\locvec^{\rm A-OED}$ and is used in the analysis step of the filter instead of $\mat{B}$.
      The resulting analysis ensemble is then integrated forward in time to the next assimilation cycle, where the procedure is repeated sequentially.

      The main difficulty of this procedure is that it requires applying a pointwise multiplication of a decorrelation matrix $\decorrmat \in \Rnum^{\Nstate \times \Nstate}$, 
      by the background error covariance matrix $\mat{B}$. As mentioned in~\S\ref{subsec:localization}, this approach is impractical in the sense that localization should not be applied directly to $\mat{B}$.

    \subsubsection{Observation-space formulation}
    \label{subsubsec:OED_localization_obs}
    A practical alternative to the state-space OED adaptive localization objective in~\eqref{eqn:B_localization_opt} can be formulated by carrying out the localization in the observation space. This alternative is discussed here.
    As described in~\S\ref{subsec:localization}, in this form of localization, $\mat{HB} $ is replaced with $\widehat{\mat{HB} } = \decorrmat^{\rm loc, 1} \odot \mat{HB} $,
    and $\mat{HB} \mat{H}\tran$ can also be replaced with $\reallywidehat{\mat{HB} \mat{H}\tran} = \decorrmat^{\rm loc, 2} \odot \mat{HB} \mat{H}\tran $.
    The localization matrix $\decorrmat^{\rm loc, 1}$ limits the effect of observations to neighboring model grid points, while $\decorrmat^{\rm loc, 2}$ localizes the correlations between various observational locations.

    Assuming that the localization radii $\locvec$ are attached to model grid points, that is, $\locvec \in \Rnum^{\Nstate}$,
    the localization matrices $\decorrmat^{\rm loc, 1} $ and $\decorrmat^{\rm loc, 2} $ can be described as
    \begin{subequations}\label{eqn:decorrelation_matrix_obs}
      \begin{eqnarray}\label{eqn:decorrelation_matrix_obs_1}
        \decorrmat^{\rm loc, 1} &=& \left[ \rho^{o|m}_{i,j} \right] \,; \quad   i=1,2,\ldots \Nobs, \text{ and } j=1,2,\ldots \Nstate \\
        \decorrmat^{\rm loc, 2} \,\equiv\, \decorrmat^{o|o} &=& \left[ \rho^{o|o}_{i,j} \right] \,; \quad i,j=1,2,\ldots \Nobs \,,
      \end{eqnarray}
    where $\rho^{o|m}_{i,j}$ is the localization coefficient~\eqref{eqn:localization_functions},
    calculated between the $i$th observation location or grid point and the $j$th model grid point.
    Similarly, $\rho^{o|o}_{i,j}$ is the localization coefficient calculated based on the distance between the $i$th  and $j$th observations.
    To calculate $\rho^{o|m}_{i,j}$, one can use the localization radius $\locrad_j$ associated with the $j$th model grid point.
    To evaluate $\rho^{o|o}_{i,j}$, however, one must attach localization radii to the observation locations.
    The simplest approach is to attach, to each observation location, the localization radius assigned
    to the closest model grid point.
    Following this approach, one can even modify $\decorrmat^{\rm loc, 1}$ to use localization radii in the observation space.
    This modification can be done by projecting the localization radii into the observation space, then calculating $\rho^{o|m}_{i,j}$ based on
    the localization radius $\decorrcoeff_i$ projected onto the $i$th observation location.
    Of course, one can modify $\decorrmat^{\rm loc, 1}$ to make the $(i,j)^{th}$ localization coefficient dependent on a localization radius $\locrad_{i,j}$.

    Following a similar argumentation as in the case of state space localization,
    one can define the localization kernel $\decorrmat^{o|o}$ in several forms.
    Here we define the observation space decorrelation kernel $\decorrmat^{o|o}$ in the form
    \begin{equation}\label{eqn:decorrelation_matrix_obs_2}
      \decorrmat^{o|o} :=
        \frac{1}{2} \left( \decorrmat^o_r + \decorrmat^o_c \right)
        = \frac{1}{2} \left( \decorrmat^o_r + \left( \decorrmat^o_r \right) \tran \right)
        = \frac{1}{2} \left[\decorrcoeff^{o|o}_{i,j} (\locrad_i) +  \decorrcoeff^{o|o}_{i,j} (\locrad_j) \right] _{i,j=1,2,\ldots,\Nobs}\,.
    \end{equation}
    \end{subequations}

    Up to this point we have been assuming that the optimization problem~\eqref{eqn:B_localization_A_opt} is solved
    to find localization radii in the model state space. Once the optimal state-space localization radii are found,
    they are projected into the observation space, for example, by using the linearized observation operator, to construct $\decorrmat^{\rm loc, 1}$
    and $\decorrmat^{\rm loc, 2}$, which are in turn used in the Kalman filter equations.
    This process reduces the cost of calculating the analysis. However, it does not help in reducing the cost of the optimization problem,
    for example, evaluating the objective and the gradient.
    Moreover, as indicated in~\cite{campbell2010vertical,flowerdew2015towards},  finding an equivalent localization in the observation space,
    given a localization in the state space, might be hard.
    
    In what follows, we assume $\locvec=(\locrad_1, \locrad_2, \ldots,\locrad_{\Nobs})\tran$ is a vector of $\Nobs$ localization radii and is used to evaluate the Kernels in~\eqref{eqn:decorrelation_matrix_obs}.
    The A-optimality criterion $\Psi^{\rm B-Loc}$ is now replaced with an alternative $\Psi^{\rm R-Loc}$ that depends on whether the localization is applied to $\mathbf{H} \mat{B}$ only or to both $\mathbf{H} \mat{B}$ and $\mathbf{HB}\mat{H}\tran$.

    \paragraph{Localizing $\mathbf{H} \mat{B}$}
      The A-optimality criterion for $\mat{R}-$localization is defined as follows:
      \begin{equation}\label{eqn:R_localization_A_opt_goal_1}
        \Psi^{\rm R-Loc}(\locvec)  
        = \Trace{  \mat{B}}
        - \Trace{
            \widehat{\mat{HB} } \,  \widehat{\mat{HB} }\tran { \left( \mat{R} + {\mat{HB} \mat{H}\tran} \right)}^{-1}
          } \,.
      \end{equation}
      As explained in \S~\ref{append:R_localization_gradient_1}, the derivative of~\eqref{eqn:R_localization_A_opt_goal_1} is
      \begin{subequations}
      \begin{equation}
        \nabla_{\locvec}{\Psi^{\rm R-Loc} }
          = -2 \, \sum_{i=1}^{\Nobs}{ \vec{e_i} \, \mat{l}_{\rm HB,i}\tran \,\psi_i  }  \,,  \\
      \end{equation}
      with $\vec{e}_i$ being the $i$th cardinality vector in $\Rnum^{\Nobs}$ and
      \begin{equation}
        \begin{aligned}
          \psi_i &= \widehat{\mat{HB} }\tran { \left( \mat{R} + {\mat{HB} \mat{H}\tran} \right)}^{-1} \, \vec{e}_i \,, \\
          \mat{l}_{\rm HB,i} &= \left(\mat{l}^s_i\right) \tran \odot \left( \vec{e}_i\tran  \mat{HB}  \right) \,, \\
          \mat{l}^s_i
            &= \left( \del{\decorrcoeff_{i,1}(\locrad_i)}{\locrad_i},
                    \del{\decorrcoeff_{i,2}(\locrad_i)}{\locrad_i},
                    \ldots,
                    \del{\decorrcoeff_{i,\Nstate}(\locrad_i)}{\locrad_i}
              \right) \tran
            \,.
        \end{aligned}
      \end{equation}
      \end{subequations}

      \paragraph{Localizing $\mathbf{H} \mat{B}$
        and $\mathbf{HB}\mathbf{H}\tran$}
      In this formulation, the OED criterion for $\mat{R}-$localization takes the form
      \begin{equation}\label{eqn:R_localization_A_opt_goal_2}
        \Psi^{\rm R-Loc}(\locvec)  
        = \Trace{ \mat{B} } -
          \Trace{  \widehat{\mat{HB}} \, \widehat{\mat{HB} }\tran { \left( \mat{R} + \reallywidehat{\mat{HB} \mat{H}\tran} \right)}^{-1}
          } \,,
      \end{equation}
      with the derivative, given the choice of $\decorrmat^{o|o}$, derived in~\S\ref{append:R_localization_gradient_2}, and is given by

      \begin{equation} \label{eqn:R_localization_derivatives}
          \nabla_{\locvec}{\Psi^{\rm R-Loc} }
            = \sum_{i=1}^{\Nobs}{
            \vec{e_i} \left( \eta^o_i - 2\, \mat{l}_{\rm HB,i}\tran \right)\, \psi^o_i }
            \,,
      \end{equation}
      with $\vec{e}_i$ being the $i$th cardinality vector in $\Rnum^{\Nobs}$, and
      \begin{subequations}\label{eqn:R_loc_strides}
        \begin{align}
          \psi^o_i &= \widehat{\mat{HB} }\tran { \left( \mat{R} + \reallywidehat{\mat{HB} \mat{H}\tran} \right)}^{-1} \, \vec{e}_i \,, \\
          \eta^o_i &= \mat{l}^o_{B,i}\, { \left( \mat{R} + \reallywidehat{\mat{HB} \mat{H}\tran} \right)}^{-1}   \widehat{\mat{HB} } \,, \\
          \mat{l}^o_{B,i} &=  \left(\mat{l}^o_i\right)\tran \odot \left({\vec{e}_i\tran \mat{HB} \mat{H}\tran}\right) \,, \\
          \mat{l}^o_i &= \left( \del{\decorrcoeff_{i,1}(\locrad_i)}{\locrad_i},
            \del{\decorrcoeff_{i,2}(\locrad_i)}{\locrad_i},
            \ldots,
            \del{\decorrcoeff_{i,\Nobs}(\locrad_i)}{\locrad_i}
            \right) \tran \,.  \label{eqn:loc_stride_vec_obs_intxt} 
       \end{align}
      \end{subequations}
      %
      %

\section{Numerical Experiments}
\label{sec:numerical_experiments}
  To study the performance of the proposed adaptive approach, we employ the two-layer Lorenz-96 model~\cite{lorenz1996predictability}.
  The numerical experiments in this work are implemented in Python
  by using \dates~\cite{attia2017dates}. 
  The source code for all experiments carried out in this work is available from~\cite{DATeS_adaptive}.

  \subsection{Experimental setup}
  \label{subsec:experimental_setup}
    Here, we describe the settings used to carry out the numerical experiments, namely, the dynamical model, the observation operator, and the metric used  to evaluate the performance of the proposed framework.

    \paragraph{Forward model: Two-layer Lorenz-96}
    We generate the truth using the two-layer model and use it to generate synthetic observations.
    Later, for the forecast step of the assimilation procedure, we employ the standard chaotic single-layer model.
    The two-layer Lorenz-96 model is given by 
    \begin{subequations}\label{eqn:two_layer_lorenz} 
    \begin{eqnarray}  
      \frac{dx_k}{dt} &=& x_{i-1} \left( x_{i+1} - x_{i-2} \right) - x_i + F - \frac{hc}{b}  \sum_{j=J(k-1)+1}^{JK}{z_j}  \,; \quad k=1,2,\ldots, K\,, \label{eqn:large_scale_lorenz} \\
        \frac{dz_j}{dt} &=& -cb\, z_{j+1} \left( z_{j+2} - z_{j-1} \right) -c z_j + \frac{hc}{b} x_{\floor{(j-1)/J}+1}   \,;    \quad j=1,2,\ldots, JK\,, \label{eqn:small_scale_lorenz}
    \end{eqnarray}
    \end{subequations}
    where all indices $j,\,k$ are periodic and $\floor{\cdot}$ is the integer floor operation.
    We set the model parameters as follows. The large-scale layer consists of $K=40$ variables while $J=32$, resulting in a small-scale layer of size $KJ=1280$.
    As in~\cite{lorenz1996predictability}, the scale ratio and coupling parametrs are set to $h=1,\, c=10,\,\text{ and } b=10$, respectively. 
    The forcing term is set to $F=8$ for consistency with the single-layer model used in the assimilation process. 
    Using the chaotic single-layer model ensures high levels of model errors during the assimilation process.
    The main objective of this particular setup is to test the accuracy as well as the robustness of the proposed approaches in the presence of model errors.
    
\begin{figure}[h]
    \centering
      \begin{subfigure}[b]{0.45\textwidth}
        \includegraphics[width=\textwidth]{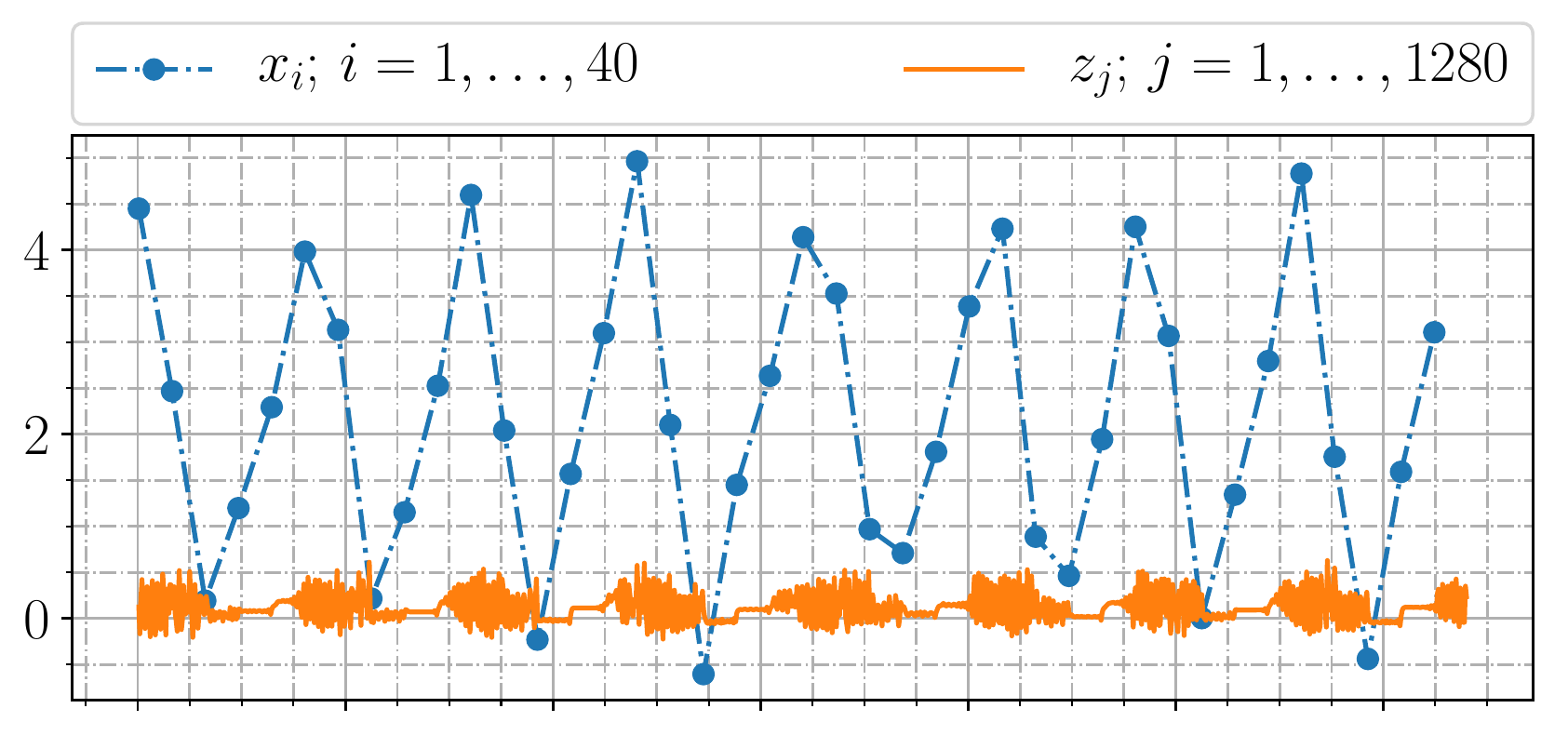}
        \caption{Reference IC}
        \label{fig:twolayer_lorenz_IC}
      \end{subfigure}
      \begin{subfigure}[b]{0.29\textwidth}
        \includegraphics[width=\textwidth]{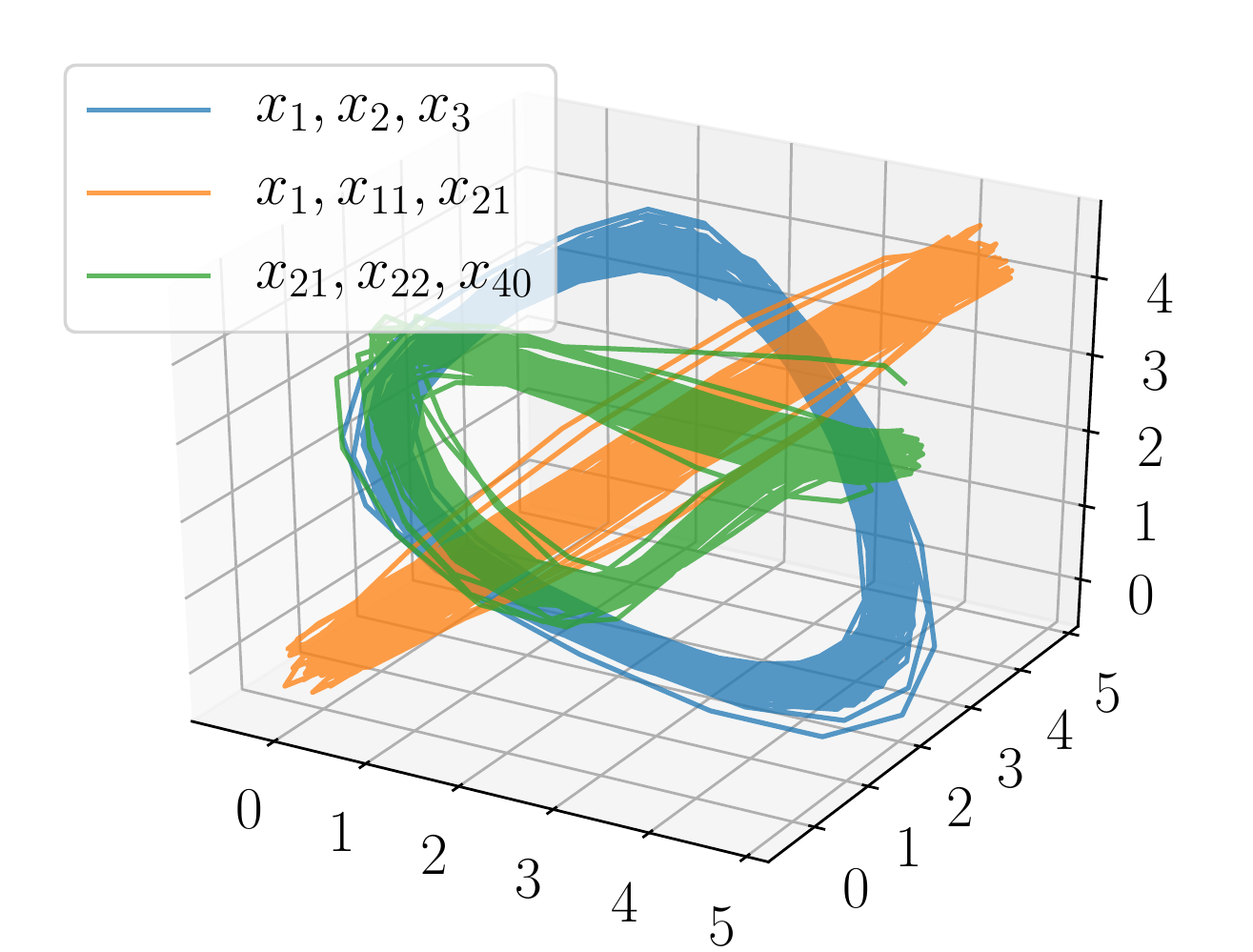}
        \caption{Reference trajectory}
        \label{fig:twolayer_lorenz_IC}
      \end{subfigure}
      \caption{The left panel (a) shows the initial state of the two-layer Lorenz-96 model~\eqref{eqn:two_layer_lorenz}. Both large-scale and small-scale components are plotted independently.
        The right panel (b) shows the true trajectory of selected components of the large-scale portion of the state vector to be reconstructed by the assimilation procedure.
        }
      \label{fig:twolayer_lorenz_Truth}
    \end{figure}

    \paragraph{Truth and synthetic observations}
    %
    We use the fourth-order explicit Runge-Kutta scheme to integrate
    the model~\eqref{eqn:two_layer_lorenz} forward in time, with a
    step size $\Delta t=0.005$ for $20,000$ time steps, resulting in an experimental timespan 
    over the time interval $[0,100]$.
    The reference trajectory, taken as the true solution, is created by forward propagation of the reference initial condition over the experiment timespan and is fixed for all experiments.
    The reference initial condition, along with the true trajectory of selected components of the state vector, is shown in Figure~\ref{fig:twolayer_lorenz_Truth}.

\begin{figure}[h]
    \centering
      \begin{subfigure}[b]{0.45\textwidth}
      \includegraphics[width=\textwidth]{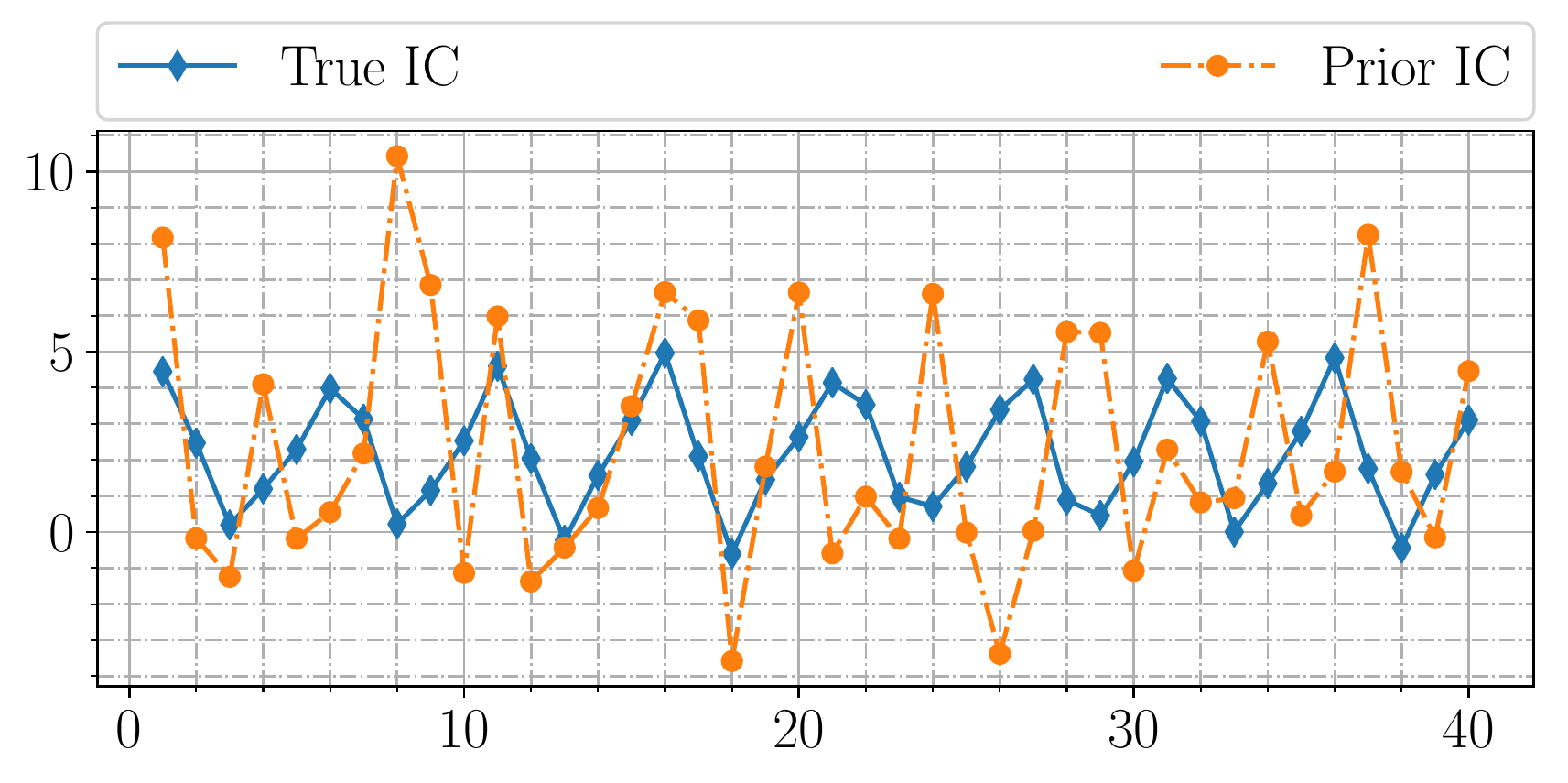}
      \end{subfigure}
      \begin{subfigure}[b]{0.45\textwidth}
      \includegraphics[width=\textwidth]{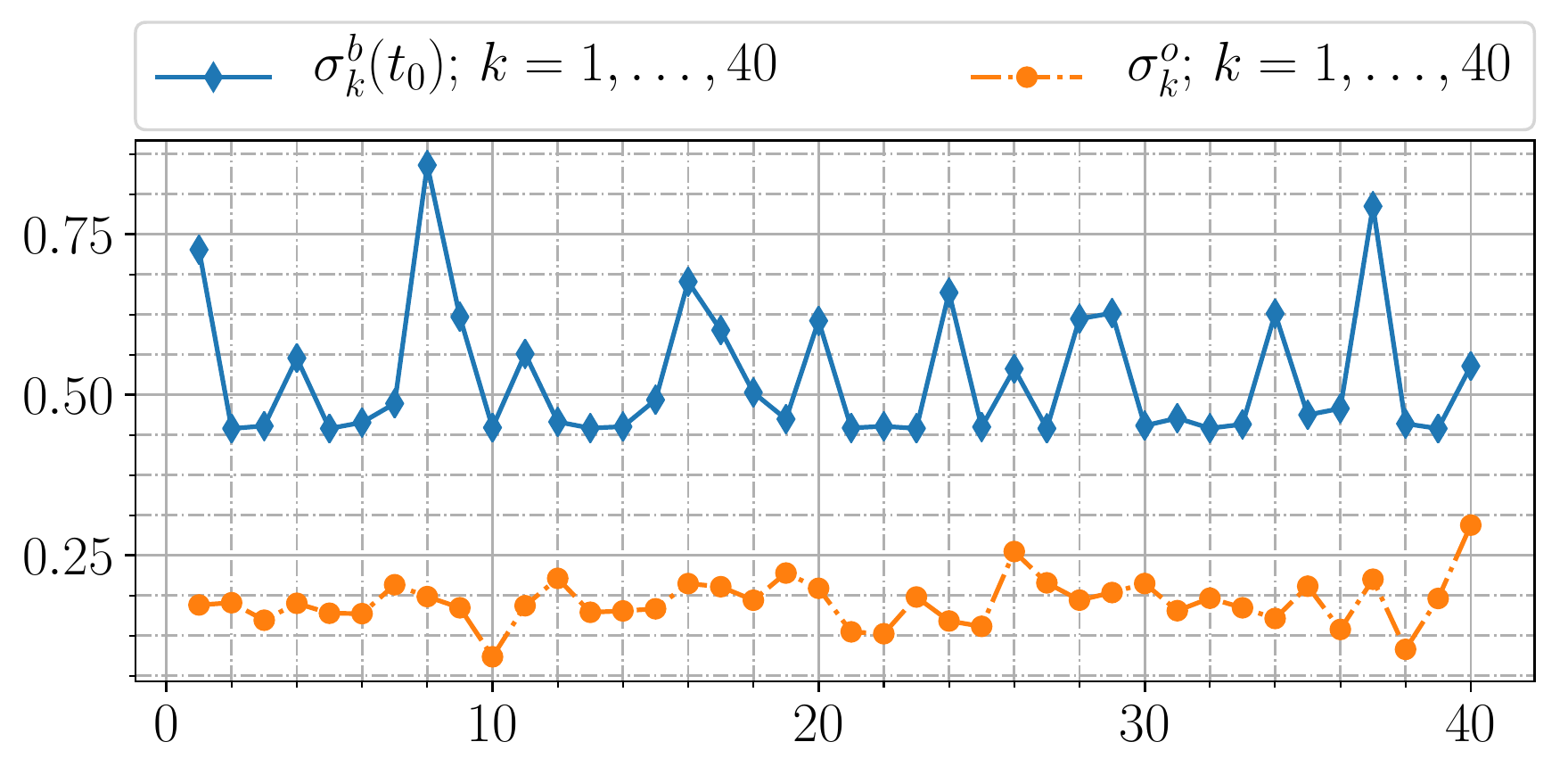}
      \end{subfigure}
      \caption{The left panel shows the prior initial condition (IC), i.e., the mean of the initial ensemble, along with the true initial condition obtained from the large-scale layer of the two-layer model~\eqref{eqn:two_layer_lorenz}.
      The right panel shows the standard deviations of the initial background distribution $\sigma^b(t_0)$  and the observation noise $\sigma^o$.
      The $x$-axis represents the entries of the state and the observation vector.
      }
      \label{fig:IC_and_noise}
    \end{figure}

    We observe the values of the large-scale variables $x_k;\,k=1,2,\ldots K=40$ every $20$ time steps, and we add random perturbations sampled from the observation noise distribution $\GM{\vec{0}}{\mat{R}}$.
    The observation error covariance matrix $\mathbf{R}$ is diagonal; in other words,  observations error are uncorrelated, with standard deviations set to $\sigma_{\rm obs} = 5\%$
    of the average magnitude of the observed reference trajectory. The observation error standard deviation, namely, the square root of the diagonal of $\mat{R}$, is plotted in Figure~\ref{fig:IC_and_noise}.
    Here, without loss of generality, we assume that the observation error covariance matrix is time independent.
    
    \paragraph{Forecast model and initial ensemble}
    In the assimilation experiments, we discard the coupled model dynamics and use only the observations generated. We use the single-layer Lorenz-96 model to produce the forecasts.
    This is equivalent to using~\eqref{eqn:large_scale_lorenz} with a coupling parameter $h=0$.
    A background  uncertainty level of $\sigma_0 = 8\%$ is used to create an initial background error covariance matrix $\mathbf{B}_0$.
    Specifically, $\mathbf{B}_0$ is created as a diagonal matrix with the variances set to $\sigma_0^2$ multiplied by the magnitude of the large-scale portion of the reference initial condition.
    The initial background state, at time $t_0=0$, is obtained by adding random noise $r_0\sim \GM{\vec{0}}{\mat{B}_0}$ to the reference initial condition.
    The initial ensemble is then created by adding i.i.d. Gaussian noise, with distribution $ \GM{\vec{0}}{\mat{B}_0}$, to the initial background state.
    The ensemble size is set to $\Nens = 25$, unless otherwise stated explicitly.
    The true solution at the initial time $t=0$, the mean of the initial prior ensemble, and the variances of the observation and initial prior error models are shown in Figure~\ref{fig:IC_and_noise}.
    %

    \paragraph{Performance metric}
      %
      The performance of ensemble DA filtering algorithms can be
      assessed by using the root mean squared error (RMSE) metric to
      monitor the  filter convergence by inspecting its ability to track the truth. It is given by
      \begin{equation}\label{eqn:RMSE_formula}
        \mathbf{RMSE} = \sqrt{\frac{1}{\Nstate} \sum_{i=1}^{\Nstate}{(x_i - x_i^{\rm True})^2} } \, ,
      \end{equation}
      where $\x\in \Rnum^{\Nstate}$ is the filter analysis (or forecast) state and $\x^{\rm True} \in \Rnum^{\Nstate}$ is the reference state (i.e., the truth)  of the system.
      The forecast RMSE is obtained by setting the $\x$ to be the mean of the forecast ensemble, and the analysis RMSE is obtained by using the analysis ensemble mean as the analysis state in~\eqref{eqn:RMSE_formula}

    \paragraph{Optimization algorithm}
      To solve the optimization problems presented in this work,
      we use SLSQP, a sequential least-squares programming subroutine~\cite{kraft1988software}. 
      The optimizer is configured with the following settings. The precision goal for the value of the objective in the stopping criterion is $ftol=10^{-6}$.
      We let the maximum number of iterations  be large,  $10,000$, to carefully test the performance and the computational cost of the approach.

    \paragraph{Benchmark settings and results}
    A desirable objective is to automatically tune the parameters in order to yield performance that is at least as good as a well-tuned version of the filter.
    Here, we  use a deterministic implementation of the EnKF (DEnKF)~\cite{sakov2008deterministic} to carry out the analysis step of the filter after solving the OED optimization problem for either inflation or localization.
    
\begin{figure}[h]
    \centering
      \begin{subfigure}[b]{0.425\textwidth}
        \includegraphics[width=\textwidth]{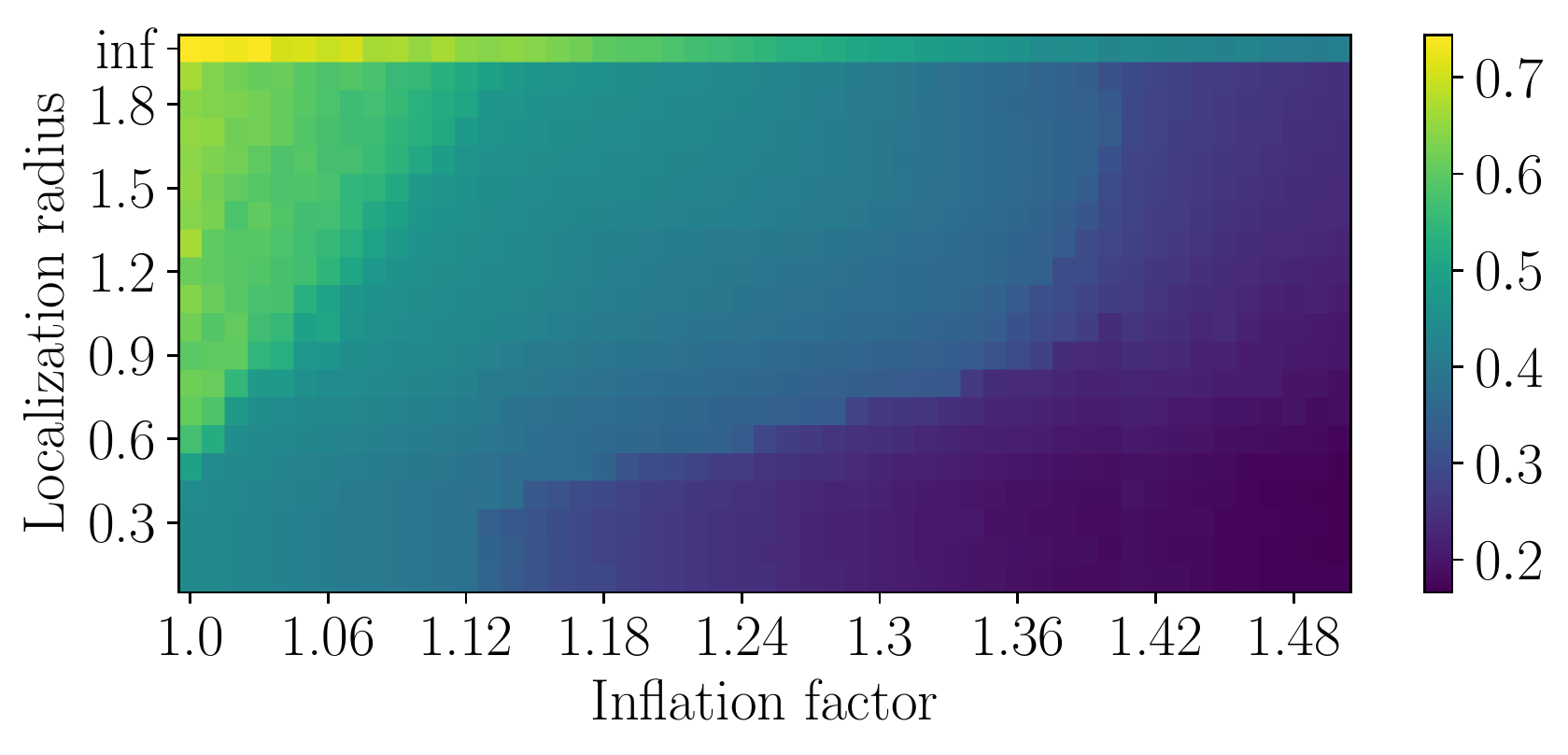}
      \end{subfigure}
      \begin{subfigure}[b]{0.45\textwidth}
        \includegraphics[width=\textwidth]{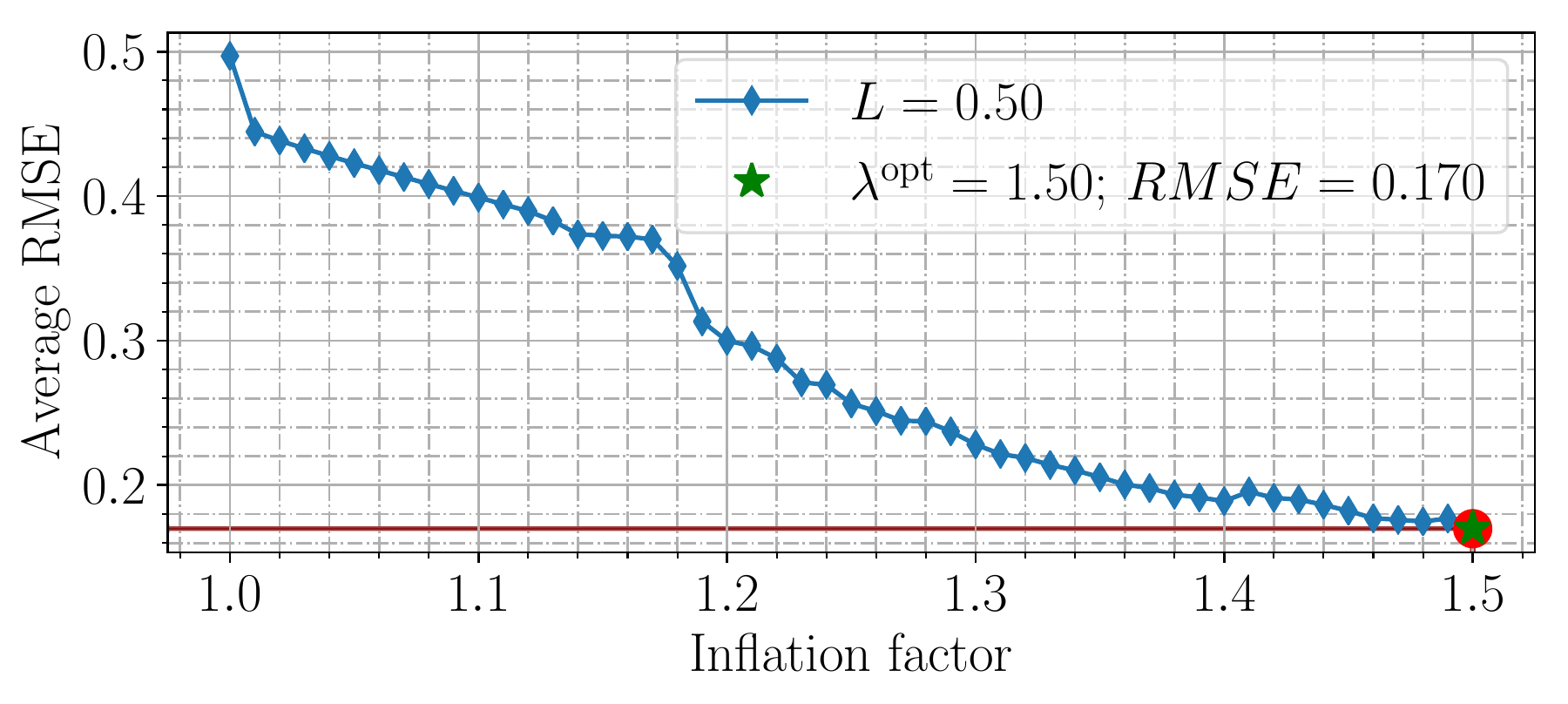}
      \end{subfigure}
      \caption{Average RMSE results obtained for various combinations of the inflation factor and localization radius.
      The RMSE is averaged over the last third of the experiment timespan, that is, over the interval $[66.7, 100]$.
      The localization function used is GC~\eqref{eqn:Gaspari_Cohn}.
      The left panel shows the RMSE results for combinations of inflation parameter and localization radius values.
      The $inf$ value of the localization radius refers to the case where no localization is carried out.
      The right panel shows the RMSE results for a specific choice of the localization radius $\locrad = 0.5$, and $51$ equally spaced values of the space-independent inflation factor over the interval $[1.0,\, 1.5]$.
      }
      \label{fig:benchmark_RMSE}
    \end{figure}

    To generate a set of benchmark results, we run experiments with various combinations of the inflation factor and the localization radius.
    Empirically tuning space-dependent inflation and localization parameters is untenable, and thus we fix them 
    over both space and time in the benchmark experiments.
    Specifically, we run the EnKF filter over the experiment timespan for combinations of $50$ equally spaced values of the inflation factor over the interval $[1, 1.5]$, along with multiple choices of the localization radius.
    Here, we report the results over the last third of the experiment
    timespan, that is, over the interval $[66.7, 100] $, to avoid spin-up artifacts. Hereafter, we use the term ``testing timespan'' to refer to this time interval.
    The RMSE results of these experiments are shown in Figure~\ref{fig:benchmark_RMSE}. 
    Results in Figure~\ref{fig:benchmark_RMSE}~(left)
    suggest that the ideal combination of parameters is located to the
    lower right corner of the plot that corresponds to a high inflation factor and a low localization radius.
    Moreover, results in~\ref{fig:benchmark_RMSE}~(right) indicate that applying more inflation in this experiment results in more accurate results.
    A combination of values corresponding to the smallest average RMSE is $\inflfac=1.5$ and $\locrad=0.5$. 
    The RMSE resulting from the experiments carried out with this choice of inflation factor and localization radius will be referred to as ``benchmark RMSE'' hereafter and will be used to judge the accuracy of the proposed methods.

  \subsection{Numerical results}
  \label{subsec:numerical_results}
    Now we show the numerical results of the OED adaptive inflation and OED adaptive localization approaches, given the numerical setup and the benchmark results described above.
    When adaptive inflation is tuned, we fix the localization radius to $\locrad=0.5$ 
    over space and time. 
    On the other hand, when adaptive covariance localization is employed, the inflation factor is fixed to $\inflfac=1.5$ over space and time.
    We notice that the behavior of both Gauss and GC localization functions in this context is similar, and consequently hereafter we assume that GC function is used.
    %

    %
    \subsection{Adaptive inflation results}
    \label{subsec:adaptive_inflation_results}
\begin{figure}[h]
      \centering
        \includegraphics[width=0.45\textwidth]{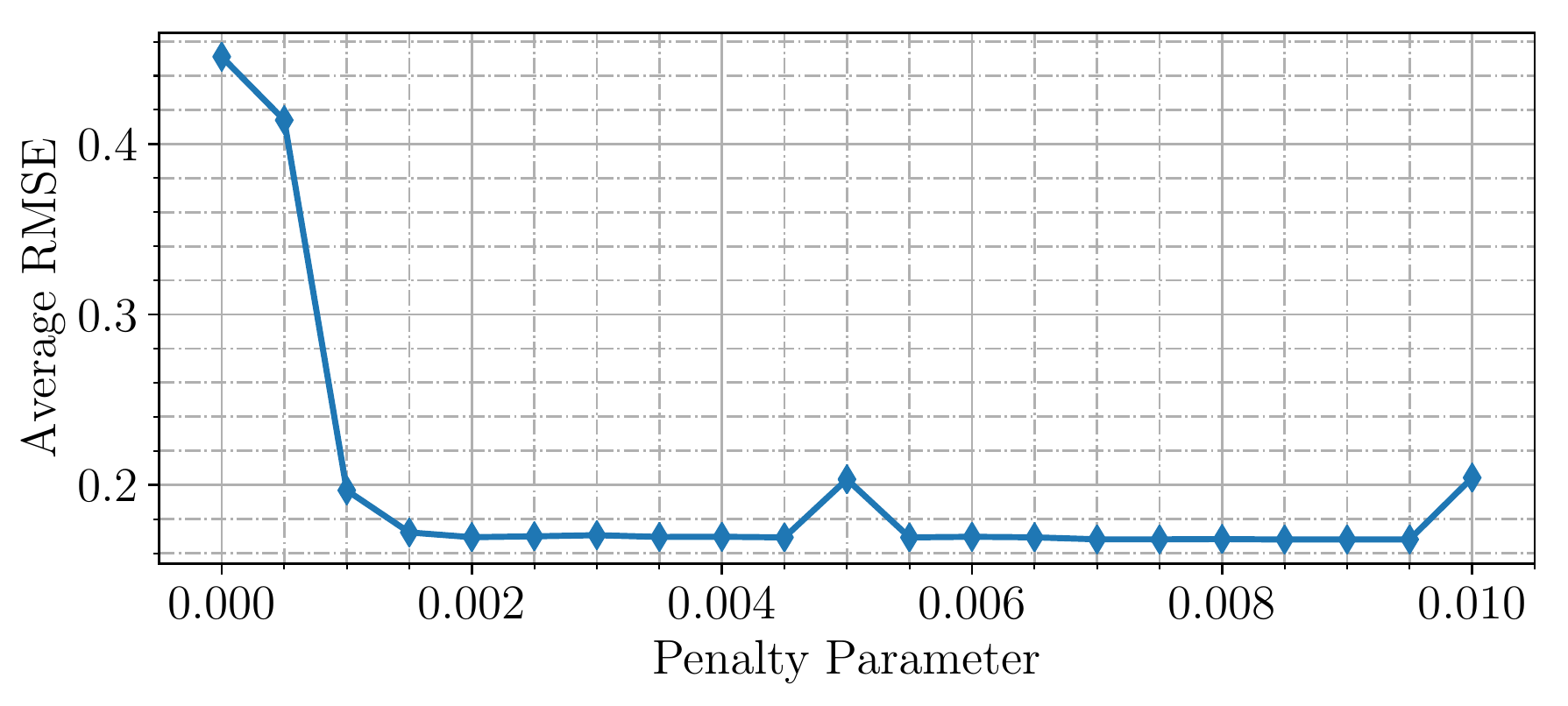}
        \caption{Average RMSE results obtained for $21$ equally spaced values of the penalty parameter $\alpha \in [0,0.01]$. 
        The RMSE is averaged over the testing timespan, that is, over the interval $[66.7, 100]$.
        }
        \label{fig:inflation_vs_avg_RMSE}
      \end{figure}

      Applying the approach presented in~\ref{subsec:OED_inflation} to automatically tune the space-time inflation factor requires choosing a proper regularization parameter $\alpha$.
      For a fair comparison with the benchmark results, here we solve the optimization problem~\eqref{eqn:multiplicative_inflation_A_opt}, where the entries of $\inflvec$ are bounded withing the interval $[1,\, 1.5]$.
      To analyze the behavior of the proposed algorithm, we show results for multiple choices of the regularization parameter $\alpha$. 
      Then we suggest an approach for choosing an appropriate value for this parameter.
      Specifically, we employ the approach described in~\ref{subsec:OED_inflation} to solve the DA filtering problem described above, for $21$ equally spaced values of $\alpha$ over the interval $[0,0.01]$.
      For each experiment, the value of $\alpha$ is fixed for all assimilation cycles.
      The average RMSE over the testing timespan is shown in Figure~\ref{fig:inflation_vs_avg_RMSE}, confirming our intuition that the penalty parameter should not be set to zero.

\begin{figure}[h]
      \centering
        \includegraphics[width=0.6\textwidth]{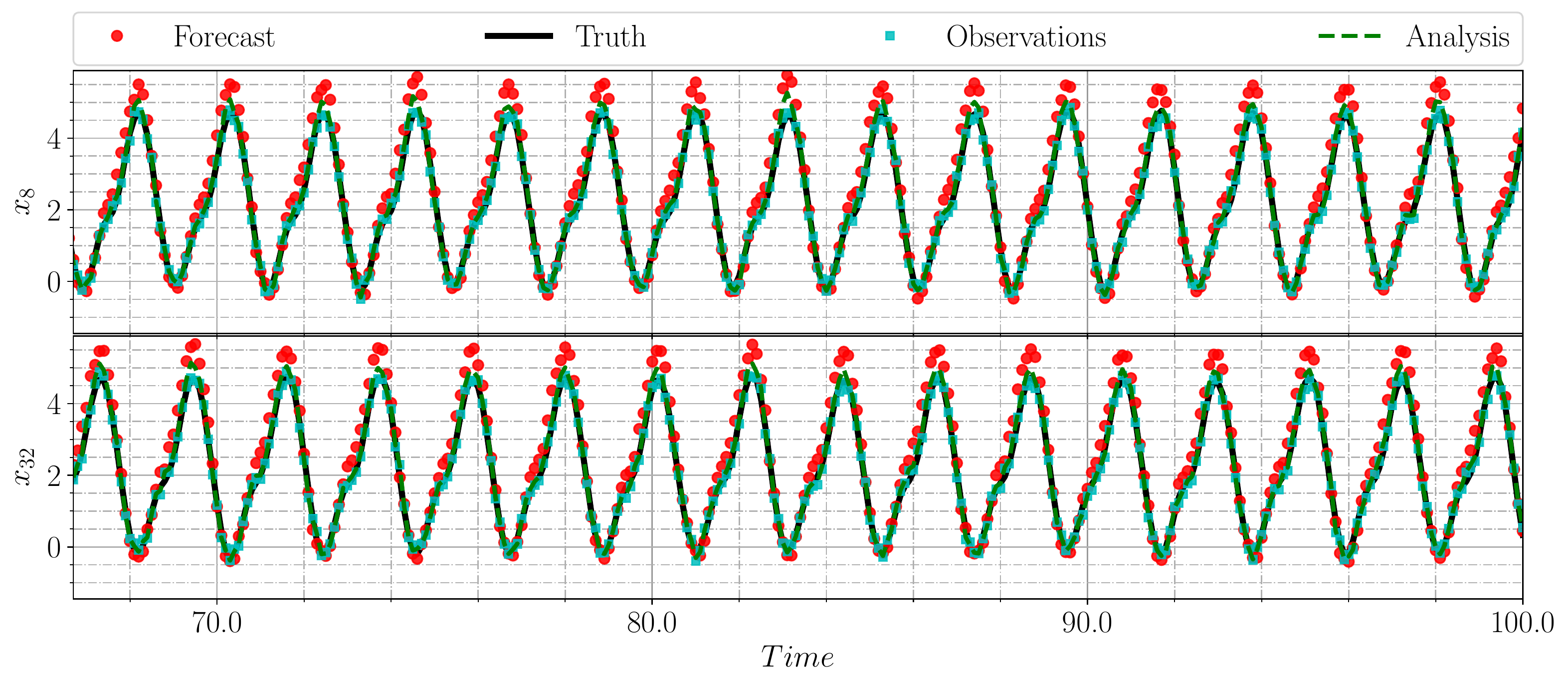}
        \caption{The truth, the observations, the forecasts, and the EnKF analysis with $\inflvec=\inflvec^{\rm A-OED}$ obtained by setting $\alpha=0.0035$.
                 The results are plotted over the testing timespan for selected entries of the state vector, specifically $x_8,\,x_{32}$.
        }
        \label{fig:adaptive_inflation_trajects}
      \end{figure}

\begin{figure}[h]
      \centering
        \includegraphics[width=0.45\textwidth]{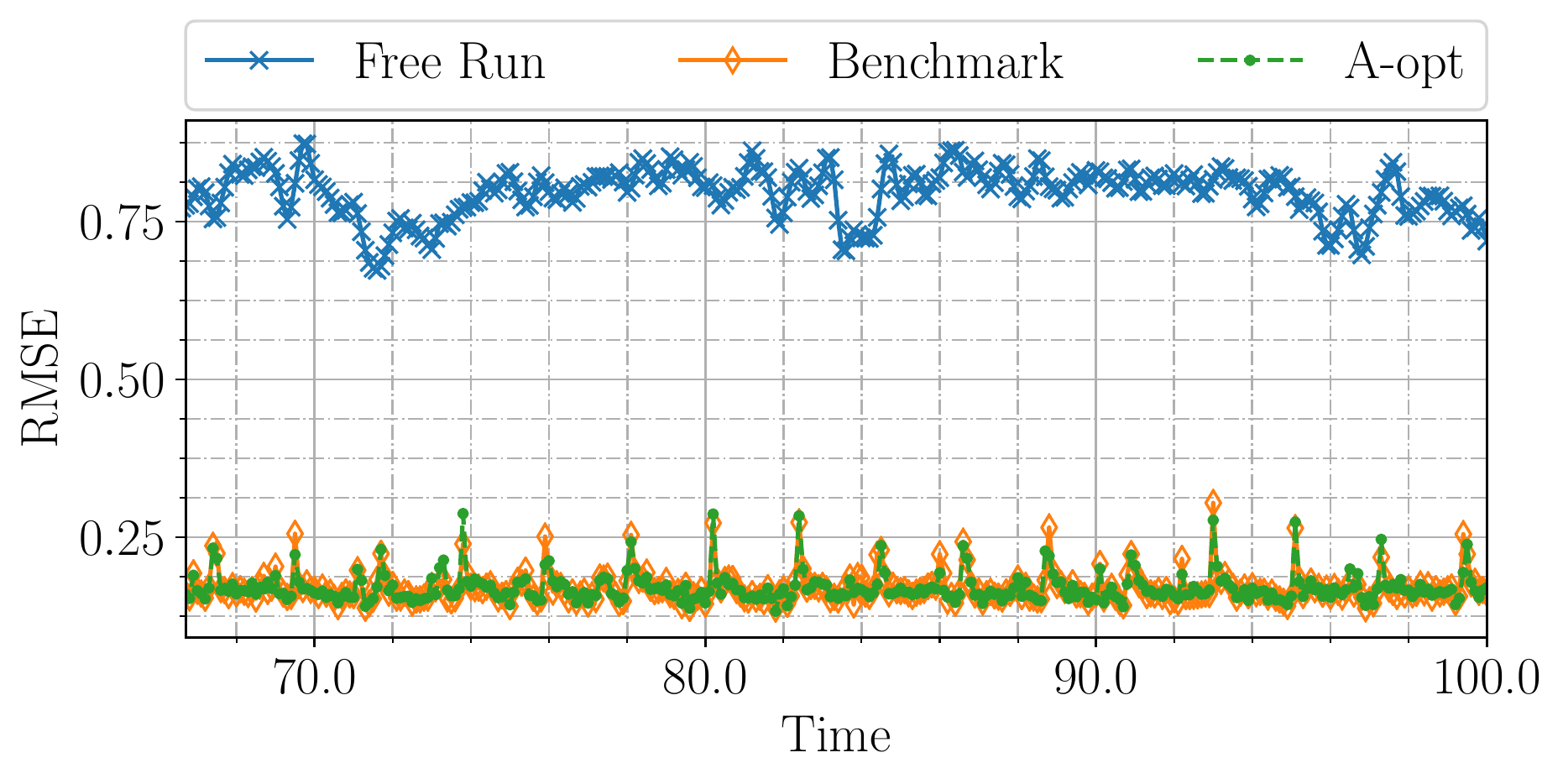}
        \caption{RMSE results over the testing timespan generated by a free run of the prior initial condition, the benchmark RMSE, and the RMSE generated by applying the OED adaptive inflation approach with $\alpha=0.0035$.
        }
        \label{fig:adaptive_inflation_RMSE_0035}
      \end{figure}

      From results in Figure~\ref{fig:inflation_vs_avg_RMSE},   we can pick a value of the penalty parameter that yields a small average RMSE, say $\alpha=0.0035$, to verify the accuracy of the filter and to study how well it tracks the truth.
      The true trajectory, synthetic observations, forecasts, and the EnKF analysis produced with the A-optimal inflation factor $\inflvec^{\rm A-OED}$ are shown in Figure~\ref{fig:adaptive_inflation_trajects}.
      The RMSE results of this experiment with OED adaptive inflation, compared with the benchmark RMSE, and the RMSE of the trajectory, from the prior initial condition, generated over the timespan without assimilation (i.e., ``free run''), are shown in Figure~\ref{fig:adaptive_inflation_RMSE_0035}.
      The results in Figures~\ref{fig:adaptive_inflation_trajects} and ~\ref{fig:adaptive_inflation_RMSE_0035} together show that, for this choice of the penalty parameter, the filter can properly estimate the true state of the underlying system.
      %

\begin{figure}[h]
        \centering
          \includegraphics[width=0.65\textwidth]{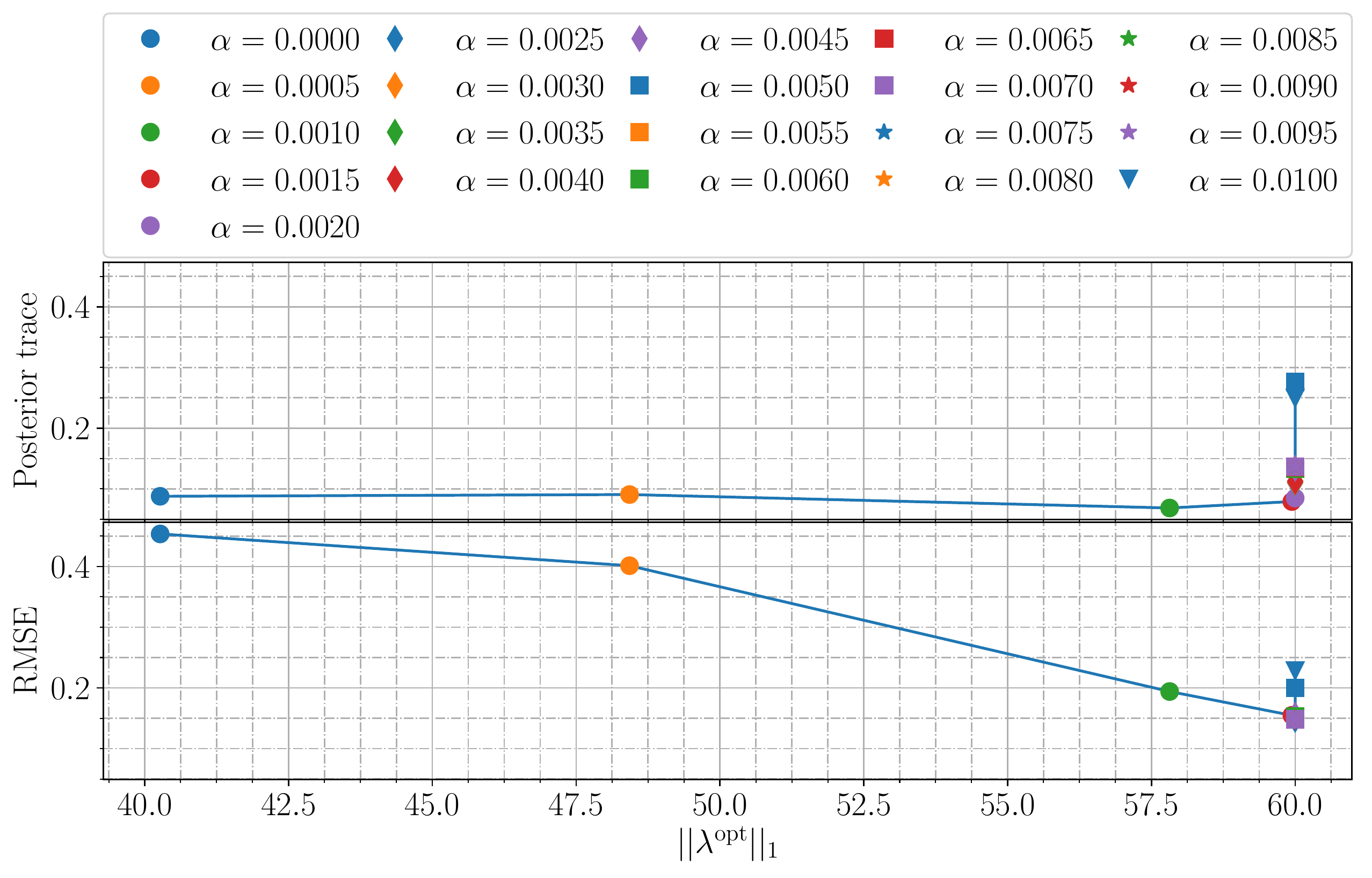}
          \caption{The top panel shows an L-curve generated at time $t=70$ for multiple A-OED adaptive inflation experiments with $21$ equally spaced values of the penalty parameter $\alpha\in[0,\,0.01]$. 
                   The analysis RMSE generated at the corresponding assimilation cycle, for each experiment, is shown in the lower panel of the figure.
          }
          \label{fig:adaptive_inflation_Lcurve}
        \end{figure}

      \paragraph{Choosing the penalty parameter $\alpha$}
      Solving the A-OED adaptive inflation problem~\eqref{eqn:multiplicative_inflation_A_opt} to find an optimal state-dependent inflation parameter $\lambda\in \Rnum^{Nstate}$ requires adjusting the value of the penalty parameter $\alpha$.
        The results in Figure~\ref{fig:inflation_vs_avg_RMSE} suggest that we should choose a value of the penalty parameter that is sufficiently greater than zero. However, increasing the penalty is expected to promote similar values of space inflation factors close to the upper bound of the optimization problem.
        Since inflation itself is an ad hoc procedure, we should target minimum inflation by choosing the smallest positive value of the penalty parameter that yields favorable RMSE. 
         In general, however, producing the results depicted in~\ref{fig:inflation_vs_avg_RMSE} is not practical.
        An alternative approach is to use an L-curve~\cite{hansen2001curve} to choose a proper value of $\alpha$. 
        Specifically, the L-curve plot is a visual tool used to choose the penalty parameter for regularization-based least squares problems.
        A typical L-curve plot draws the norm of the optimal solution on one axis, against the value of the objective on the other axis, for multiple choices of the regularization parameter.
        The optimal penalty occurs at the elbow of the curve, that is, the point of maximum curvature.
        Since we are solving the OED problem at every assimilation cycle, we choose a time point, for example $t=70$, and generate the L-curve from the results of the assimilation cycle at this time instant.
        Figure~\ref{fig:adaptive_inflation_Lcurve} shows the L-curve with the norm of the optimal inflation factor $\inflvec^{\rm A-OED}$ on the x-axis, against the trace of the posterior ensemble covariance matrix on the y-axis.
        Note that the curve here is mirrored compared with the standard L-curve plot because of the sign of the regularization term.
        The RMSE results are also added to the same plot, to inspect how the filter behaves for increasing values of the penalty parameter.
        We note that all the assimilation cycles yielded similar plots.

\begin{figure}[h]
        \centering
          \includegraphics[width=0.45\textwidth]{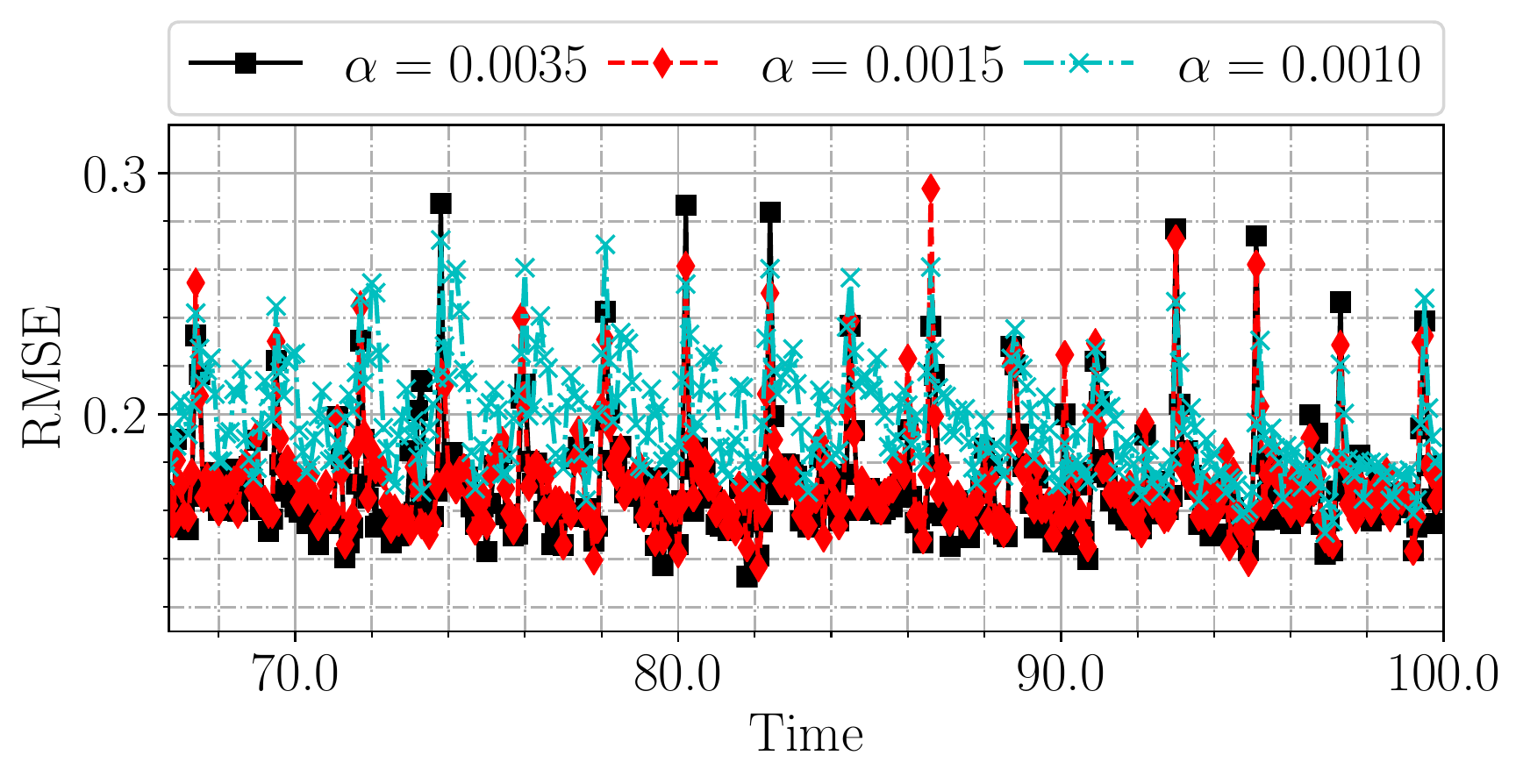}
          \caption{RMSE results over the testing timespan generated by applying the OED adaptive inflation approach with penalty parameter $\alpha$ set to $0.0035,\, 0.0015,\, \text{ and } 0.0010$, respectively.
          }
          \label{fig:adaptive_inflation_RMSE_001_0015_0035}
        \end{figure}

\begin{figure}[h]
        \centering
          \begin{subfigure}[b]{0.32\textwidth}
          \includegraphics[width=\textwidth]{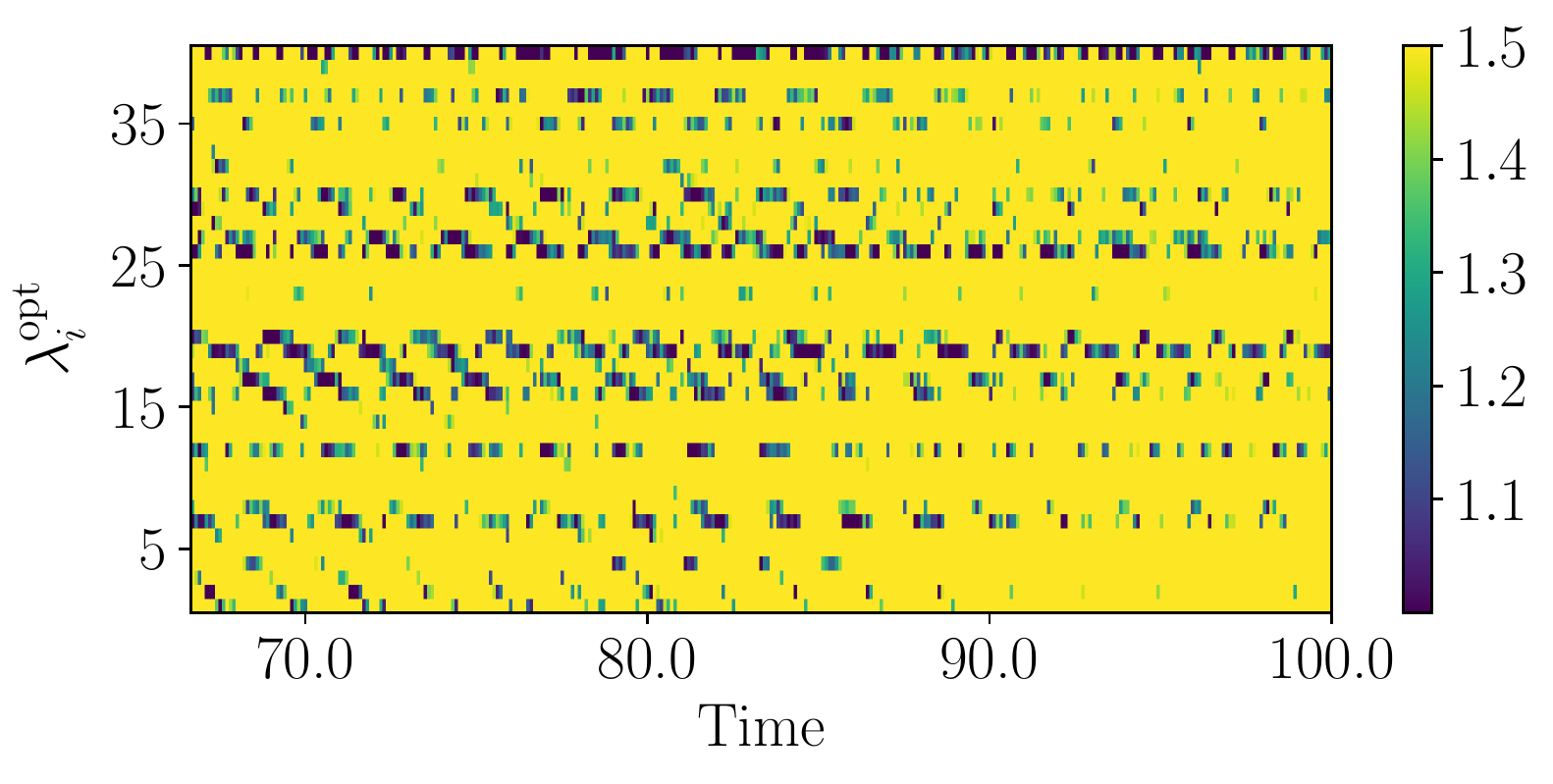}
            \caption{$\alpha=0.0010$}
            \label{fig:adaptive_inflation_spacetime_001}
          \end{subfigure}
          \begin{subfigure}[b]{0.32\textwidth}
          \includegraphics[width=\textwidth]{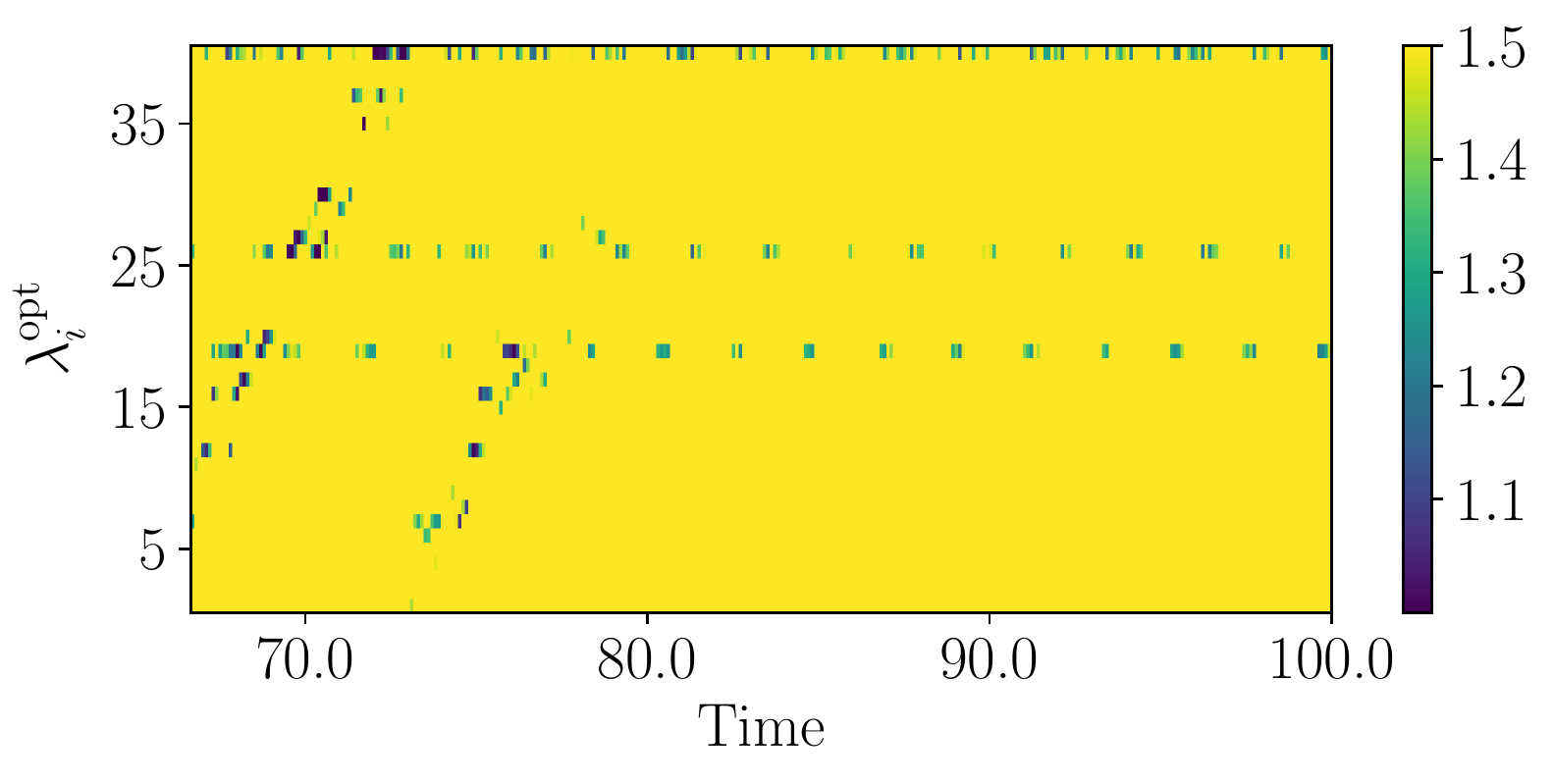}
            \caption{$\alpha=0.0015$}
            \label{fig:adaptive_inflation_spacetime_0015}
          \end{subfigure}
          \begin{subfigure}[b]{0.32\textwidth}
          \includegraphics[width=\textwidth]{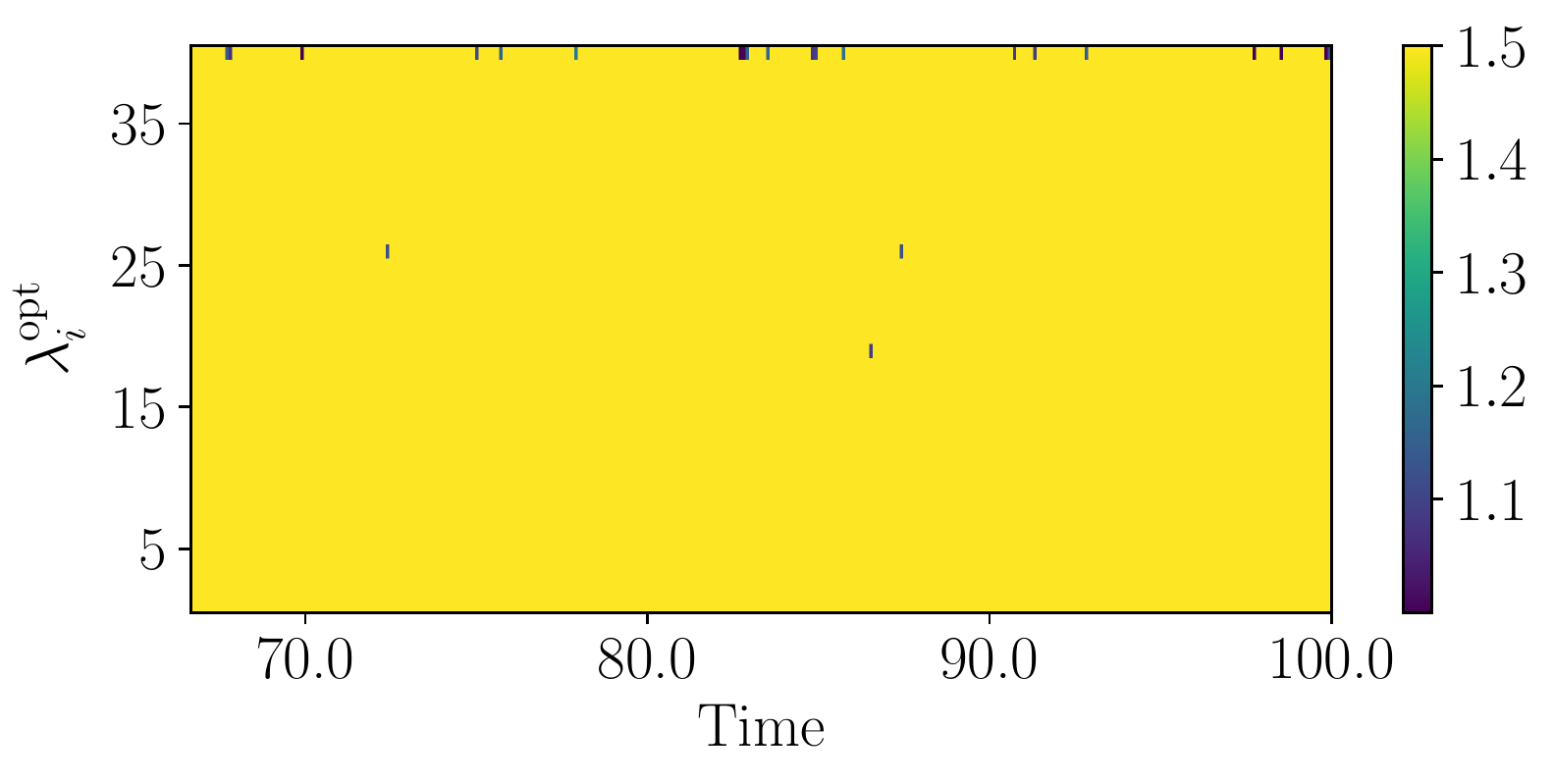}
            \caption{$\alpha=0.0035$}
            \label{fig:adaptive_inflation_spacetime_0035}
          \end{subfigure}
          \caption{Space-time evolution of the A-optimal inflation factor $\inflvec^{\rm A-OED}$, over the testing timespan,  with penalty parameter $\alpha$ set to $0.0010,\, 0.0015,\, \text{ and } 0.0035$, respectively.
          }
          \label{fig:adaptive_inflation_spacetime}
        \end{figure}

\begin{figure}[h]
        \centering
          \includegraphics[width=0.80\textwidth]{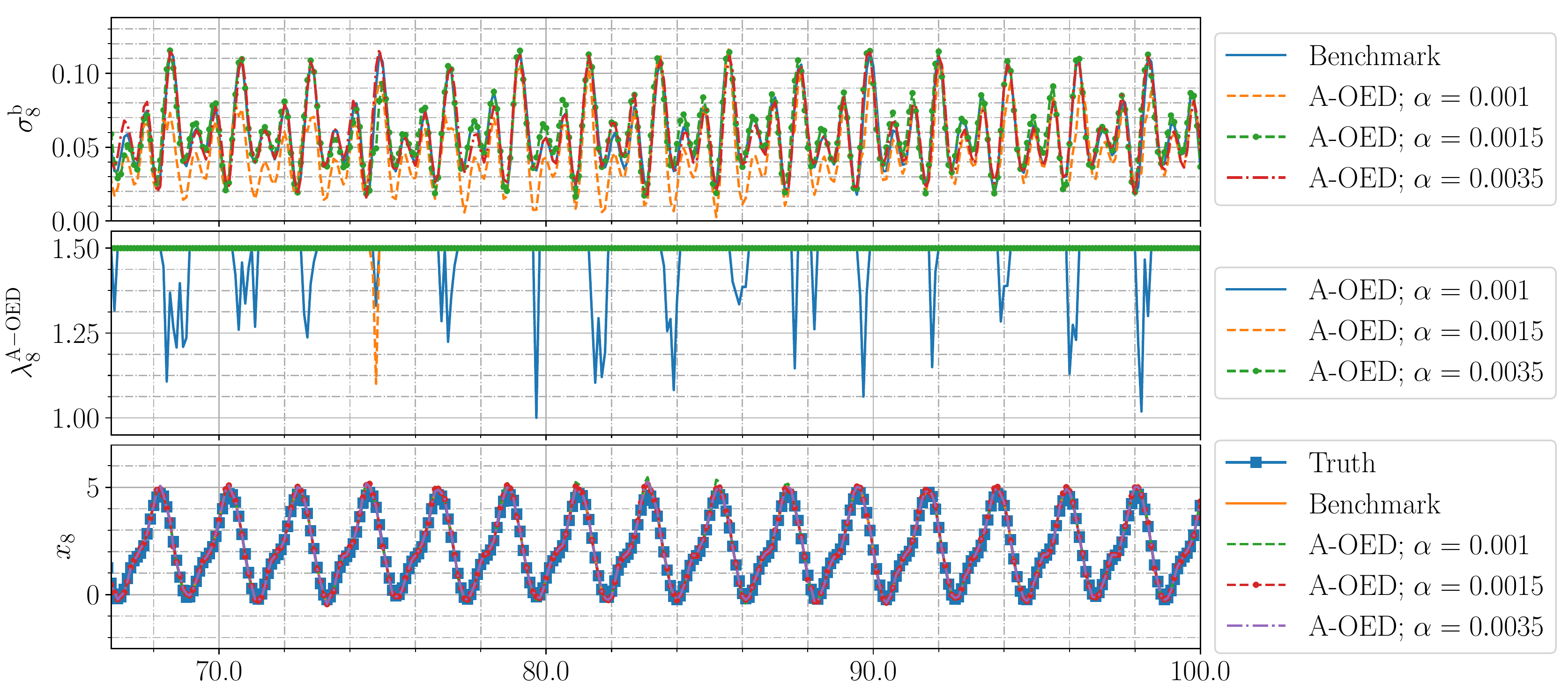}
          \caption{Data assimilation results over the testing timespan of the benchmark compared with the truth, and with the OED adaptive space-time inflation approach with multiple choices of the regularization parameter $\alpha$.
          Results are shown for a specific entry of the model state $x_8$ and the corresponding A-OED inflation factor $\inflfac_8^{\rm A-OED}$ with $\alpha=0.001,\, 0.0015,\, 0.0035$, respectively.
          The top panel shows the ensemble-based standard deviation of $x_8$ for both the benchmark and the adaptive inflation experiments.
          The middle panel shows the value of the optimal inflation factor $\inflfac_8^{\rm A-OED}$, and the lower panel shows the recovered state $x_8$ compared with the truth.
          }
          \label{fig:adaptive_infl_traject_and_noise}
        \end{figure}

        These results suggest that by increasing the penalty parameter, the RMSE decreases while the posterior trace is almost similar until the elbow of the L-curve is hit.
        By further increasing the value of the penalty parameter, the
        RMSE may increase,  amplifying the chance that the filter might diverge,
        especially if the bound constraints in the optimization problem are relaxed.
        An ideal choice of the penalty parameter is a value around $\alpha=0.0015$ that resides at or close to the elbow of the L-curve.
        The RMSE results for experiments carried out with $\alpha=0.0010$ and  $\alpha=0.0015$, compared with the earlier choice of $\alpha=0.0035$, are shown in Figure~\ref{fig:adaptive_inflation_RMSE_001_0015_0035}.
        While these results reveal similar RMES behavior, the main difference between these three choices, as suggested earlier, is the amount of inflation exerted. 
        This difference is further explained by the space-time evolution of the inflation factor shown in Figure~\ref{fig:adaptive_inflation_spacetime}, 
        which indicates that increasing the penalty parameter promotes higher levels of inflation.
        These results are also supported by the results in Figure~\ref{fig:adaptive_infl_traject_and_noise}~(top), which shows the standard deviation of $x_8$, the $8$th entry of the state vector calculated from the forecast ensemble for the benchmark experiment as well as the adaptive inflation experiments. By increasing the value of the penalty parameter, the inflation factors tend to increase, and consequently the variability of state variables increases.
        Figure~\ref{fig:adaptive_infl_traject_and_noise}~(middle) shows the behavior of the inflation factor that applies to $x_8$ over time, and Figure~\ref{fig:adaptive_infl_traject_and_noise}~(bottom) shows the corresponding entry of the analysis state obtained from the benchmark experiment and the adaptive inflation experiments compared with the truth. These results explain the capability of the approach to tune the space-dependent inflation factors that enable the filter to produce accurate forecasts.

\begin{figure}[h]
        \centering
          \includegraphics[width=0.40\textwidth]{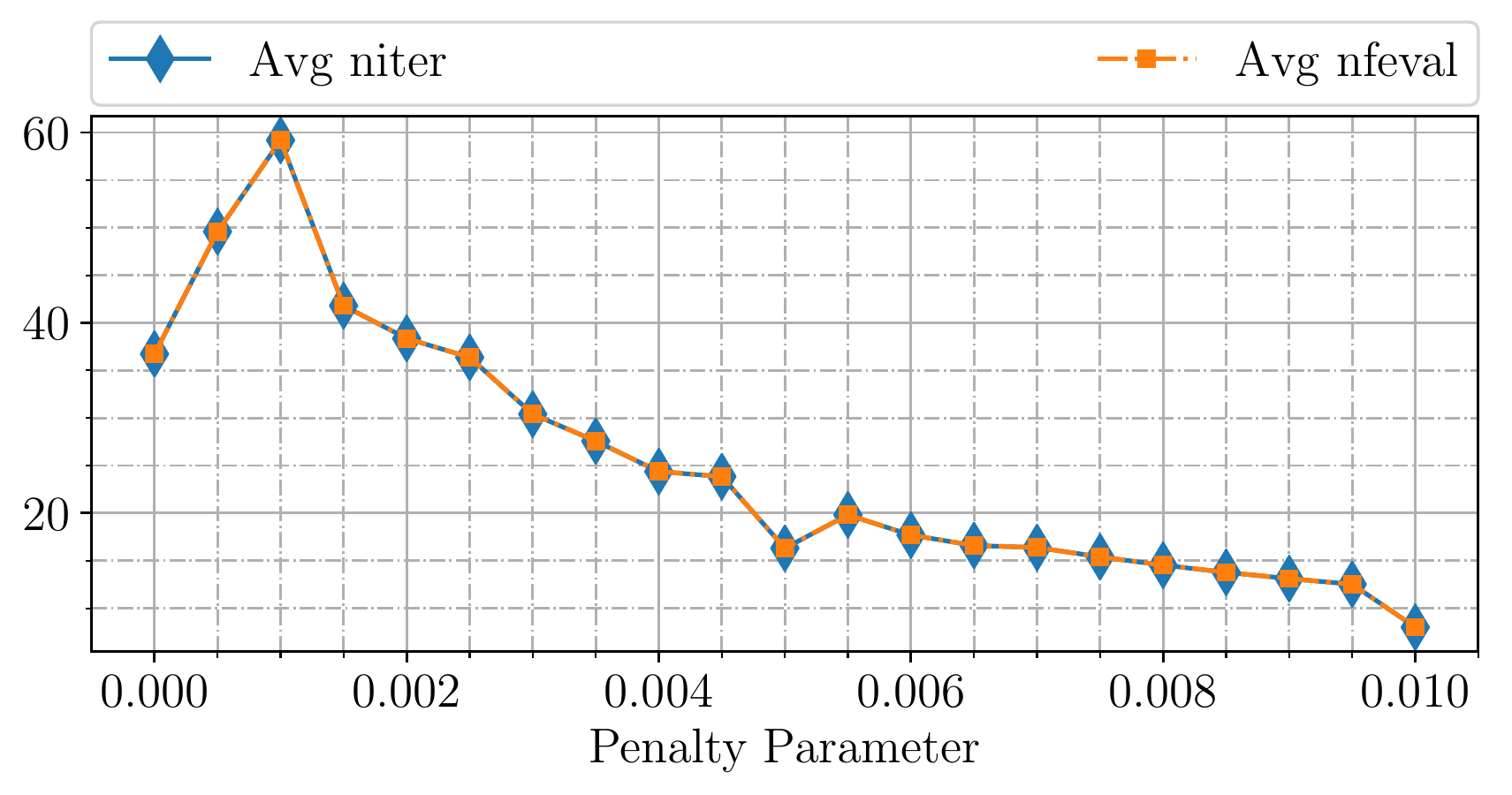}
          \caption{Average number of iterations ``\textit{Avg niter}'' and number of function evaluations ``\textit{Avg nfeval}'' of the optimization procedure for values of the penalty parameter in the interval $[0,\, 0.01]$.
          }
          \label{fig:adaptive_inflation_niter}
        \end{figure}

        Another effect of changing the penalty parameter is the number of iterations required by the optimization procedure to find an optimal solution of the A-optimality problem~\eqref{eqn:multiplicative_inflation_A_opt}.
        The number of iterations averaged over the assimilation cycles in the testing timespan for various values of the penalty parameter $\alpha$ is shown in Figure~\ref{fig:adaptive_inflation_niter}.
        These results suggest that the number of iterations spent in the optimization procedure decreases by increasing the penalty parameter.
        %

    %
    \subsection{Adaptive localization results}
    \label{subsec:adaptive_localization_results}
\begin{figure}[h]
      \centering
        \begin{subfigure}[b]{0.40\textwidth}
        \includegraphics[width=\textwidth]{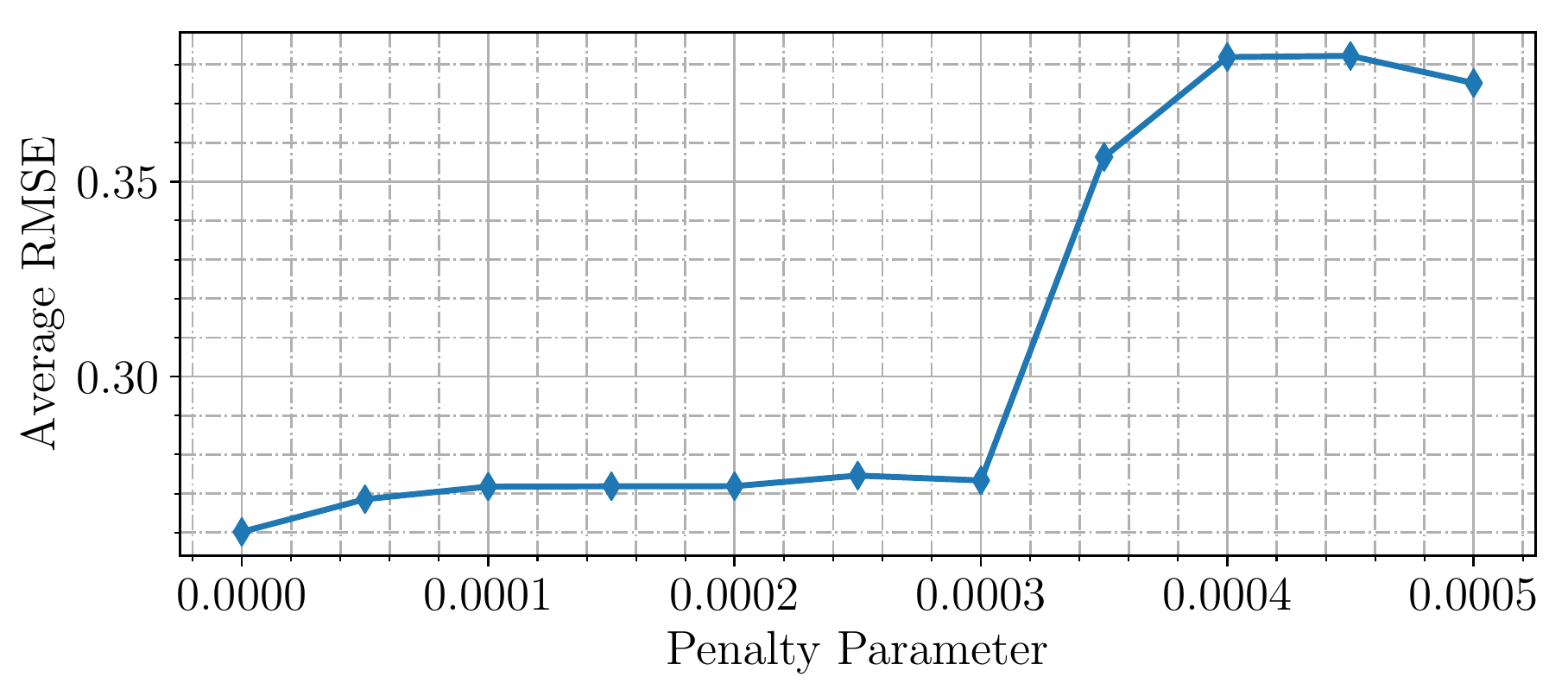}
        \end{subfigure}
        \begin{subfigure}[b]{0.40\textwidth}
        \includegraphics[width=\textwidth]{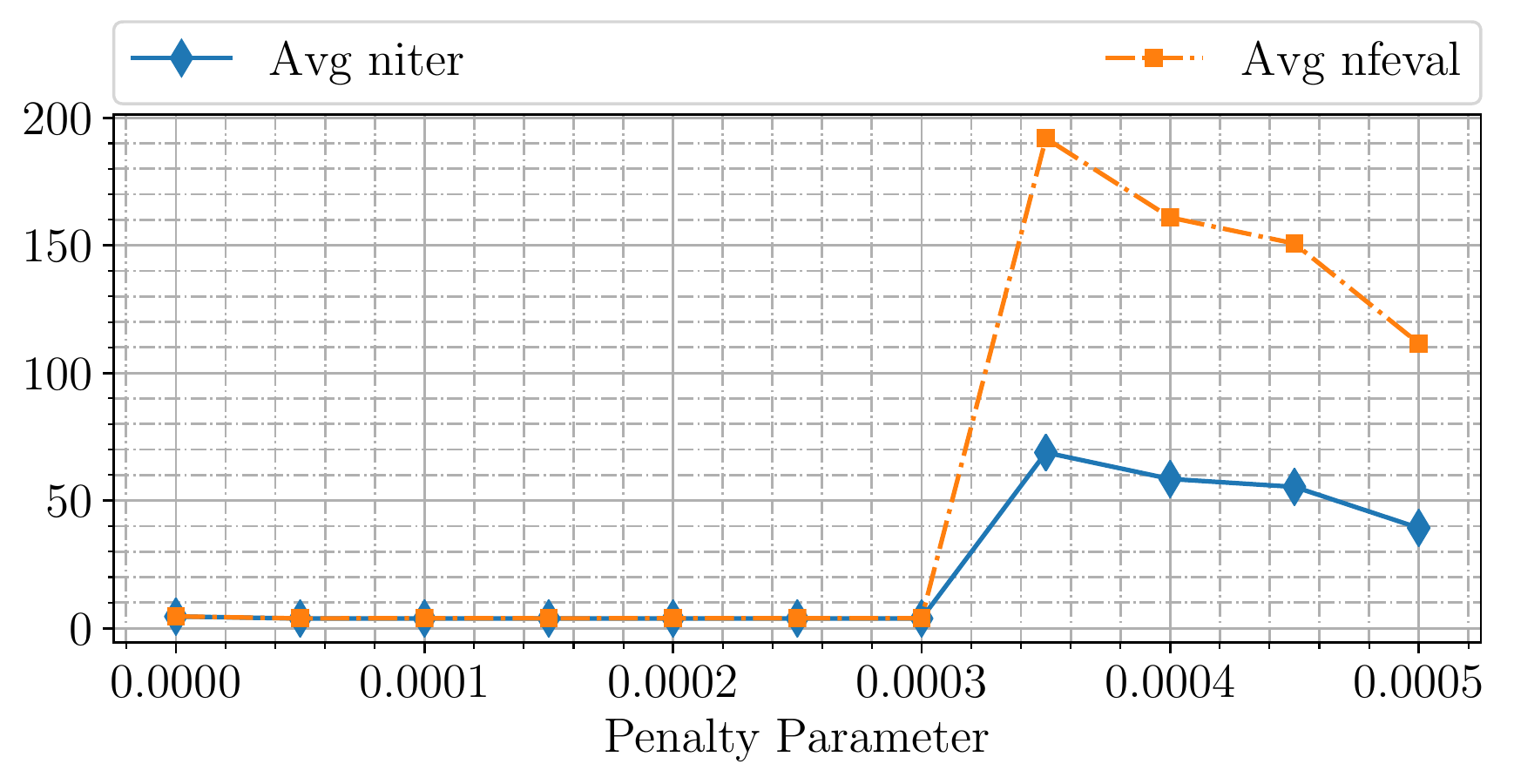}
        \end{subfigure}
        \caption{Average RMSE results obtained for $21$ equally spaced values of  the penalty parameter $\gamma \in [0,0.01]$.
                 The RMSE is averaged over the last third of the testing timespan, that is, over the interval $[66.7, 100]$. 
        }
        \label{fig:adaptive_localization_avg_RMSE_and_niter}
      \end{figure}

      %
      In this section we discuss the results of the adaptive
      space-time covariance localization approach introduced in~\S\ref{subsec:OED_localization}.
      Figure~\ref{fig:adaptive_localization_avg_RMSE_and_niter} shows
      the average RMSE and the number of optimization iterations of the EnKF experiments with A-OED adaptive localization for various values of the penalty parameter $\gamma$.
      Similar to the case with adaptive inflation, for every choice of the penalty parameter we run a DEnKF experiment, where~\eqref{eqn:B_localization_A_opt} is solved at every assimilation cycle
      and the optimal solution, namely, $\locvec=\locvec^{\rm A-OED}$, is used  in the localization step of the filter.
      The results in Figure~\ref{fig:adaptive_localization_avg_RMSE_and_niter}~(left) suggest that the penalty term in~\eqref{eqn:B_localization_A_opt}
      is not necessarily required to achieve good performance of the
      adaptive filter. 
      Moreover, as indicated by the results in 
      Figure~\ref{fig:adaptive_localization_avg_RMSE_and_niter}~(right),
      increasing the penalty parameter $\gamma$ is more likely to increase the computational cost.
      Forcing regularization by increasing the penalty parameter $\gamma$ restricts the values of the localization radii as well as its variability and  can result in higher computational cost and even degradation of the filter accuracy.
      
\begin{figure}[h]
      \centering
        \includegraphics[width=0.80\textwidth]{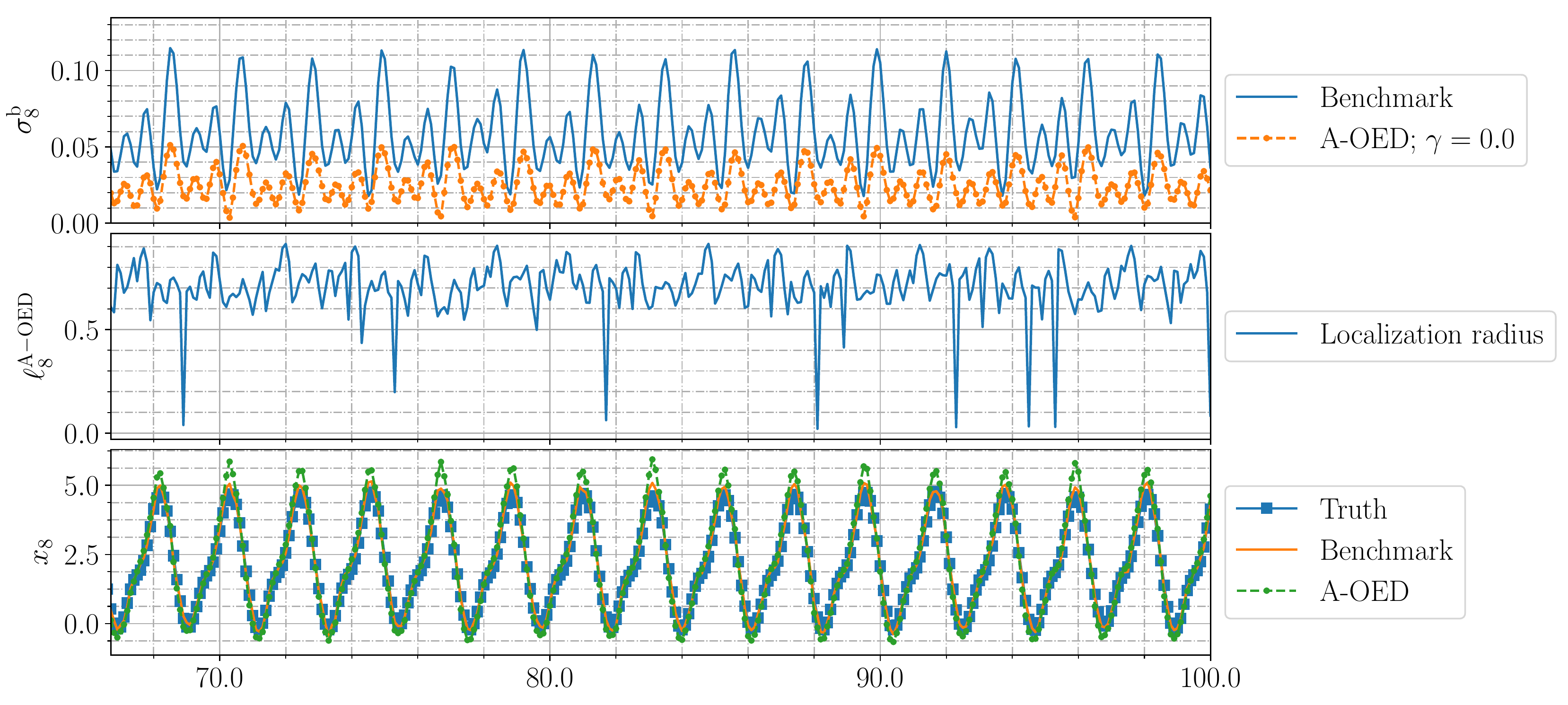}
        \caption{Data assimilation results, over the testing timespan, of the benchmark results compared with the truth, and the OED adaptive space-time localization with $\gamma=0$.
        Results are shown for a specific entry of the model state $x_8$ and the corresponding A-OED localization radius $\locrad_8^{\rm A-OED}$. 
        The top panel shows the ensemble-based standard deviation of $x_8$ for both the benchmark and the adaptive localization experiments.
        The middle panel shows the value of the A-optimal localization radius $\locrad_8^{\rm A-OED}$, and the lower panel shows the recovered state $x_8$.
        }
        \label{fig:adaptive_loc_traject_and_noise}
      \end{figure}

      We show results of the adaptive space-time localization experiments for $\gamma=0$. Similar to the results in Figure~\ref{fig:adaptive_infl_traject_and_noise}, 
      in Figure~\ref{fig:adaptive_loc_traject_and_noise} we show the results from a specific entry of the state vector and the optimal localization radius over time.
      The top panel shows the standard deviation of the $8$th entry of the state vector $x_8$, calculated from the forecast ensemble for both the benchmark experiment and the adaptive localization experiment. 
      The middle panel shows the $8$th entry of the localization radius vector $\locrad_8^{\rm A-OED}$, and the lower panel shows the recovered corresponding entry of the model state. 
      These results show the ability of the OED adaptive approach to properly tune the space-dependent localization radii and hence enable the filter to recover the truth. 
      %

\begin{figure}[h]
      \centering
        \includegraphics[width=0.70\textwidth]{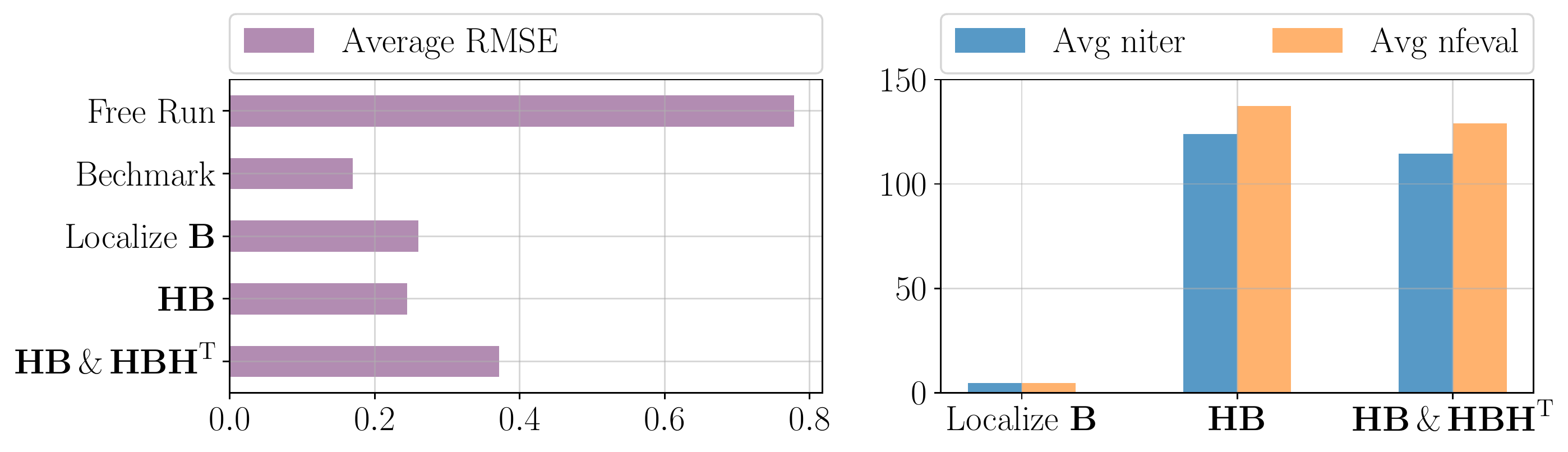}
        \caption{OED adaptive covariance localization results. The average RMSE of the benchmark experiment compared with the free run and the OED adaptive localization is shown in the left panel of the figure.
        The OED adaptive localization is solved in the state space with the optimality criterion~\eqref{eqn:B_localization_objective} and in the observation space with criterion~\eqref{eqn:R_localization_A_opt_goal_1} and  with criterion~\eqref{eqn:R_localization_A_opt_goal_2}.
        The average number of iterations of the optimization procedure in each case is shown in the right panel of the figure.
        }
        \label{fig:B_R_localization_bars}
      \end{figure}

    \paragraph{Localization in the observation space}
      We now show the results of the adaptive covariance localization approach, where the OED objective and its associated gradient are approximated by projection into the observation space.
      Note that in the numerical experiments carried out in this work, the dimension of the observation space is the same as that of the model state space. 
      This means that the only differences in this approach involve the way the localization is carried out while calculating the objective function and the associated gradient.
      Figure~\ref{fig:B_R_localization_bars} shows the RMSE results and
      the computational cost resulting from using the criteria
      \eqref{eqn:B_localization_objective},~\eqref{eqn:R_localization_A_opt_goal_1},
      and \eqref{eqn:R_localization_A_opt_goal_2}, respectively. 
      While the RMSE is slightly lower in the case of calculating the objective by applying localization to $\mat{HB}$ only, the three cases show acceptable tracking of the truth. 
      However, the computational cost in the cases where the objective is calculated by applying localization in the observation space is much higher. Specifically, the number of iteration spent by the optimizer to converge to an optimal localization radius is much higher that the case of state space formulation.
      The main reason is that the value of the objective
      function at many iterations becomes negative. Since the objective
      value at least approximates the value of the posterior trace, it must not be
       allowed to take a negative value. For this reason, we added an
       additional constraint to the optimizer $\Psi^{\rm Infl}>0$
       that  resulted in much higher cost. 
       While the idea of projecting the OED objective function and the gradient into the observation space is suggested in order to reduce the computational cost and memory requirements, the numerical results shown here indicate that it can produce negative effects on the optimization procedure. This idea is still worth being considered further in future work, however, in order to promote the feasibility of the OED adaptive covariance localization to real large-scale applications.
      %

      \subsection{Robustness of the approach} \label{subsec:robustness}
\begin{figure}[h]
        \centering
          \begin{subfigure}[b]{0.45\textwidth}
          \includegraphics[width=\textwidth]{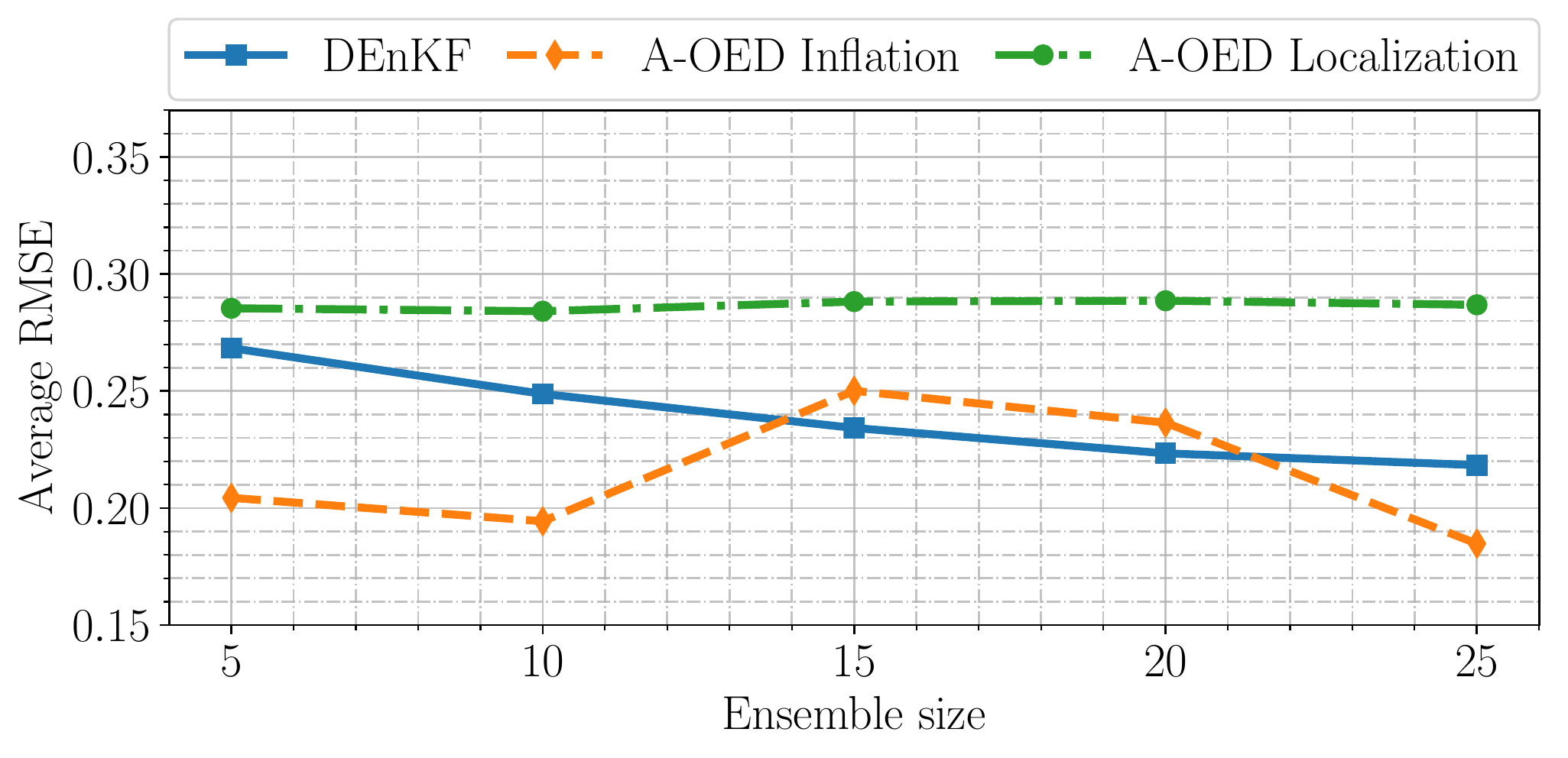}
          \end{subfigure}
          \begin{subfigure}[b]{0.45\textwidth}
          \includegraphics[width=\textwidth]{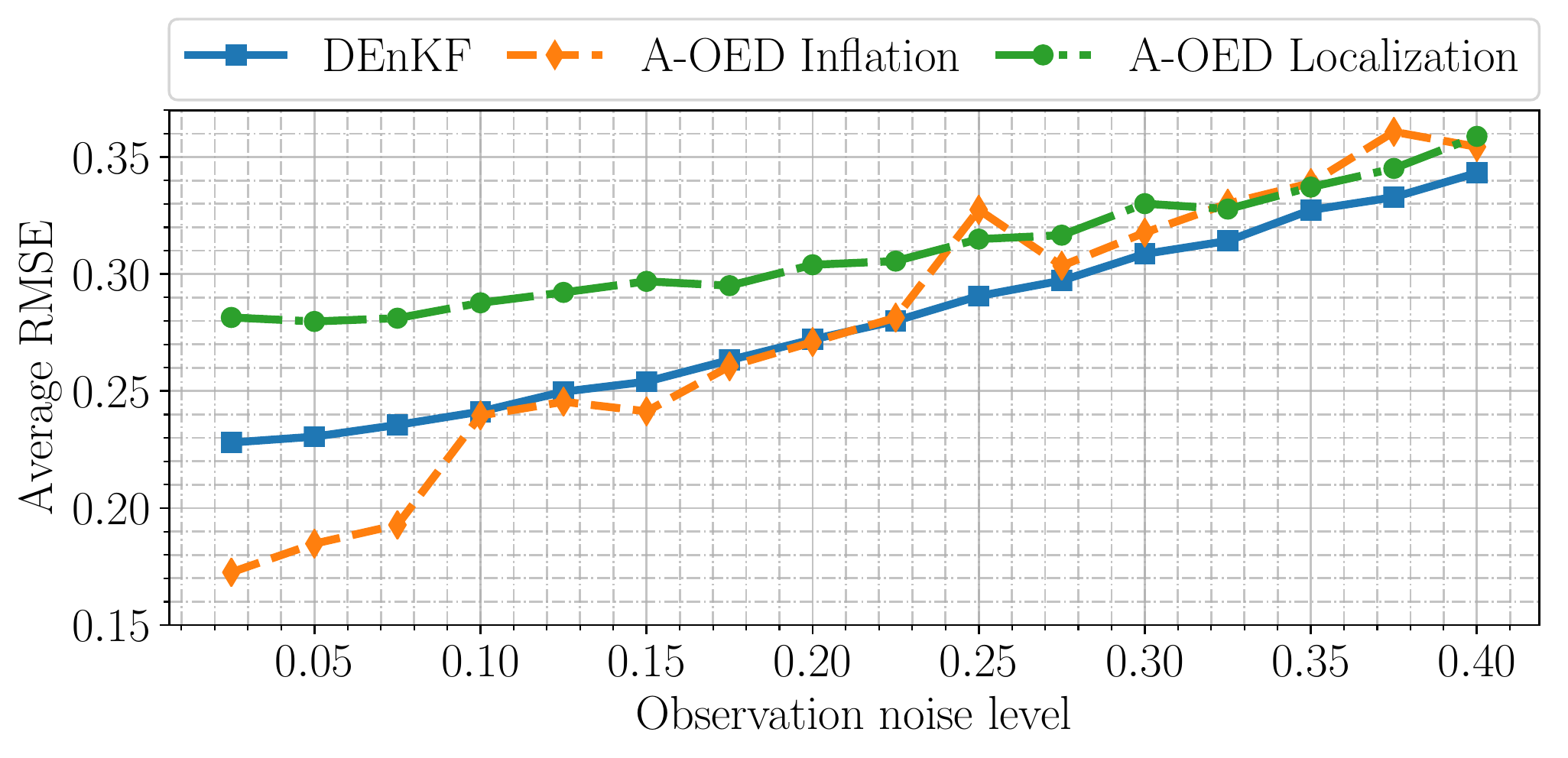}
          \end{subfigure}
          \caption{RMSE results for experiments carried out with various ensemble sizes and observation noise levels. The OED adaptive inflation problem is solved with penalty parameter $\alpha=0.0015$.
          The left panel shows the RMSE averaged over the testing timespan for assimilation experiments carried out with ensemble sizes $\Nens=5, 10, 15, 20, 25$.
          The right panel show the average RMSE for experiments carried out with increasing observation noise level from $2.5\%$ to $40\%$. 
          }
          \label{fig:adaptive_inflation_VarEns_and_Noise}
        \end{figure}

        To test the robustness of the proposed approach, we study the behavior for multiple choices of the ensemble size and various levels of observation noise.
        We compare the RMSE results of adaptive inflation and adaptive localization with the results of a standard DEnKF with inflation and localization taken from the benchmark experiment and  fixed over both time and space.
        The RMSE results in Figure~\ref{fig:adaptive_inflation_VarEns_and_Noise}~(left) show the RMSE averaged over the testing timespan for assimilation experiments carried out with ensemble sizes $\Nens=5, 10, 15, 20, 25$. 
        Figure~\ref{fig:adaptive_inflation_VarEns_and_Noise}~(right) shows the average RMSE for experiments carried out with the noise level ranging from $2.5\%$ to $40\%$.
        These results confirm the robustness of the proposed adaptive inflation approach against changing the ensemble size and the observation error.
        %

\section{Discussion and Concluding Remarks}
\label{sec:conclusions}
  %
  Ensemble data assimilation algorithms follow a Monte Carlo approach to quantify the uncertainty associated with the model state.
  Using a finite ensemble of model states is known to cause underestimation of state covariances and is mainly responsible for development of long-range spurious correlations.
  Covariance underestimation drives the filtering algorithm away from
  the observations by overtrusting the model forecast, which leads to filter divergence.
  The ensemble-based covariance matrix is generally rank-deficient and must be corrected by removing spurious correlations.
  Covariance underestimation is alleviated by applying covariance inflation, where the deviations of ensemble members from their mean are magnified by an inflation factor.
  Spurious correlations are removed by applying covariance localization, where the correlations are damped with distance within a localization radius of influence.
  Both the inflation factor and the localization radius can be varied with space and time to enhance the performance of the assimilation system.
  Tuning these parameters (i.e., inflation factor and localization radius) is a critical task that is usually carried out by trial and error and is computationally expensive to apply.

  In this study, we developed a new framework for adaptive tuning of the space-time inflation factor and localization radius of influence.
  The optimal parameter is defined as the one that minimizes the posterior uncertainty in the model state.
  We define an objective for each of these parameters independently, where the solution is regularized by using a predefined norm.
  The objective function, defined as the trace of the filtering posterior covariance matrix, is
  known in the optimal experimental design community as the A-optimal criterion.
  The main goal of the proposed approach is to tune the space-time
  inflation or the localization parameters without the need for
  trial-and-error tuning.
  The framework presented in this study does not assume independence of observation or background errors; hence, it is useful for dense as well as sparse observational networks.
  Moreover, the algorithms can in principle be applied without much expert knowledge about the model or the observations.
  This claim, however, will be further tested operationally in future studies.
  Numerical experiments to test the proposed algorithm were carried out by using the coupled two-layer Lorenz-96 model with a deterministic ensemble Kalman filter.
  The performance of the filter with optimal adaptive inflation factors and optimal adaptive localization radii was independently compared with the filtering results obtained with empirically well-tuned parameters and with the truth.

  While carrying out adaptive inflation or adaptive localization by solving the A-optimality criterion is promising, one has to be careful when applying both simultaneously.
  Here, we formulate the objective function as the function that minimizes the posterior uncertainty given that only either inflation or localization is applied;
  we simply update one of these parameters at every assimilation
  cycle, while keeping the other fixed.
  Alternatively,  the A-optimal objective function can be reformulated, given that both inflation and localization are applied.
  One approach for calculating the optimal inflation factor and localization radii simultaneously is to formulate a weighted objective
  consisting of the posterior traces obtained by inflation and localization independently.
  These extensions are beyond the scope of this study and will be explored in the future.

  Covariance localization can be carried out in the model state space or the observation space.
  In order to apply the covariance localization, a decorrelation kernel must be specified. The decorrelation kernel is pointwise multiplied by the ensemble background error covariance matrix
  or its projection into the observation space.
  In this work, we report specific formulation of the decorrelation kernel and use it for formulating the objective functions and gradients.
  However, the approach presented here can be easily modified to handle other decorrelation kernels.
  Unlike the decorrelation kernel discussed here, in ideal settings a decorrelation kernel should be symmetric positive definite; however, this is difficult to achieve in practice.
  Moreover, while the result of the pointwise product of two positive definite matrices is also positive definite, the inverse does not necessarily hold.
  Hence, even if the decorrelation kernel is not positive definite, the localized flow-dependent background covariance matrix is not necessarily indefinite.
  Moreover, we are more interested in preserving positive definiteness of the posterior covariance matrix rather than the prior covariance.
  While the numerical results carried out in this study do not reveal any urgent need for a positive definite kernel, this issue deserves further investigation.
  %

  Although the forecast ensemble mean is not affected by inflation or localization, the posterior moments are highly influenced.
  Spatially and temporally varying inflation factors and localization radii are tuned to counter the loss of variance and the developed spurious correlations;
  however, they can result in physical imbalance by destroying smoothness of the assimilated states.
  The simplest approach to handle this issue is to enforce smoothness on the solution of the adaptive inflation or localization optimization problems.
  This process can be carried out by using regularization or by simply applying a moving average to the optimal solution.
  Another approach is to use any probabilistic knowledge about the tuning parameter, instead of regularization. We plan to investigate this approach in future work.

  Minimizing the trace of the posterior covariance is equivalent to minimization of the expected squared errors in the linear framework.
  The numerical results presented in this work provide compelling evidence of the merits of our approach in the linear setup.
  Natural extensions of the proposed framework are to consider nonlinear observation operators and to use probabilistic knowledge about the tuning parameter as a form of regularization
  leading to a probabilistic version of the developed framework.

\appendix

\section{OED Adaptive Inflation Gradient}
\label{append:A_inflation_gradient}
  Here, we derive the gradient of the A-optimality criterion for adaptive space-inflation.
  The derivative of the objective~\eqref{eqn:Infl_A_Opt_Objective} is derived as follows. For each $i=1,2,\ldots,\Nstate$
  \begin{equation} \label{eqn:infl_objective_grad_1}
    \begin{aligned}
      \del{ \Psi^{\rm Infl} (\inflvec) }{\inflfac_i} 
      = \Trace{\del{\widetilde{\mat{A}}} {\inflfac_i}  }
      %
      &= - \Trace{ \widetilde{\mat{A}}
      \del{\left( \mat{D}^{-\frac{1}{2}}  \mat{B}^{-1} \mat{D}^{-\frac{1}{2}} + \mat{H}\tran \mat{R}^{-1} \mat{H} \right)}{\inflfac_i}
      \widetilde{\mat{A}}
      } 
      = - \Trace{ \widetilde{\mat{A}} \del{\left( \mat{D}^{-\frac{1}{2}}  \mat{B}^{-1} \mat{D}^{-\frac{1}{2}} \right)}{\inflfac_i}  \widetilde{\mat{A}} } \,.
    \end{aligned}
  \end{equation}

  Note that the entries of the inflated prior covariance matrix relate to the original ensemble covariance matrix via the relation
  $ \left[ \mat{D}^{\frac{1}{2}} \mat{B} \mat{D}^{\frac{1}{2}} \right] _{i,j} =  \sqrt{\inflfac_i \inflfac_j} \left[ \mat{B} \right]_{i,j} $.
  This means that the derivative of the inflated matrix with respect to $\inflfac_i$ is a symmetric sparse matrix with zeros everywhere except
  the $i$th row and $i$th column such that the $j$th entry of each of these rows and columns is $\frac{1}{2}\sqrt{\frac{\inflfac_j}{\inflfac_i}}$,
  where $i\neq j$, and is $1$ for $i = j$.
  By definition, $\mat{D}^{-\frac{1}{2}} = \sum_{i=1}^{\Nstate}{ \inflfac_i^{-1/2}  \vec{e}_i \vec{e}_i\tran}$, and consequently
  $ \del{}{\inflfac_i} {\mat{D}^{-\frac{1}{2}} } = - \frac{1}{2} \inflfac_i^{-3/2}  \vec{e}_i \vec{e}_i \tran $. By utilizing this fact and using
  the product rule of derivatives, it follows that
  \begin{equation}\label{eqn:infl_objective_grad_2}
    \begin{aligned}
      \del{\left( \mat{D}^{-\frac{1}{2}}  \mat{B}^{-1} \mat{D}^{-\frac{1}{2}} \right)}{\inflfac_i}
      & 
      = \frac{- \inflfac_i^{-3/2} }{2} \left(\vec{e}_i \vec{e}_i \tran \mat{B}^{-1} \mat{D}^{-\frac{1}{2}} + \mat{D}^{-\frac{1}{2}} \mat{B}^{-1} \vec{e}_i \vec{e}_i \tran \right) \,,
    \end{aligned}
  \end{equation}
  where $\vec{e}_i$ is the $i$th cardinality vector in $\Rnum^{\Nstate}$.
  By employing the fact that the matrix trace is a linear mapping and by
  substituting~\eqref{eqn:infl_objective_grad_2} in~\eqref{eqn:infl_objective_grad_1}, we obtain the following:
  \begin{equation} \label{eqn:infl_objective_grad_3}
    \begin{aligned}
      \del{ \Psi^{\rm Infl} (\inflvec) }{\inflfac_i} 
      &= \frac{ \inflfac_i^{-3/2}}{2} \, \left(
      \Trace{ \widetilde{\mat{A}}  \vec{e}_i \vec{e}_i \tran   \mat{B}^{-1} \mat{D}^{-\frac{1}{2}} \widetilde{\mat{A}} }
      + \Trace{ \widetilde{\mat{A}}  \mat{D}^{-\frac{1}{2}}  \mat{B}^{-1}   \vec{e}_i \vec{e}_i \tran \widetilde{\mat{A}} }
      \right)  \\
      &= \inflfac_i^{-3/2} \, \Trace{ \vec{e}_i \tran   \mat{B}^{-1} \mat{D}^{-\frac{1}{2}}  \widetilde{\mat{A}}  \widetilde{\mat{A}}   \vec{e}_i }
      = \inflfac_i^{-3/2} \, \vec{e}_i \tran   \mat{B}^{-1} \mat{D}^{-\frac{1}{2}}  \widetilde{\mat{A}}  \widetilde{\mat{A}}   \vec{e}_i
      = \inflfac_i^{-1} \,  \vec{e}_i \tran \widetilde{\mat{B}}^{-1}  \widetilde{\mat{A}} \widetilde{\mat{A}}   \vec{e}_i  \,,
    \end{aligned}
  \end{equation}
  where the last step makes use of the fact that $\mat{D}^{\frac{1}{2}}$ is a diagonal matrix with $\sqrt{\inflfac_i}$ on the diagonal,
  and consequently $\vec{e}_i \tran \mat{D}^{\frac{1}{2}} = \sqrt{\inflfac_i}\, \vec{e}_i\tran$.
  We can use  the following identity:
  \begin{equation}\label{eqn:infl_post_identity}
      \widetilde{\mat{B}}^{-1}  \widetilde{\mat{A}}
      = \widetilde{\mat{B}}^{-1}  \left( \widetilde{\mat{B}}^{-1} + \mat{H}\tran \mat{R}^{-1} \mat{H} \right)^{-1}
      = \left( \mat{I} + \mat{H}\tran \mat{R}^{-1} \mat{H} \widetilde{\mat{B}}  \right)^{-1}
      = \mat{I} - \mat{H}\tran \left( \mat{R} + \mat{H} \widetilde{\mat{B}} \mat{H}\tran  \right)^{-1}  \mat{H}  \widetilde{\mat{B}}  \,.
  \end{equation}
  From~(\ref{eqn:infl_objective_grad_3},~\ref{eqn:infl_post_identity}), it follows immediately that
  \begin{equation}\label{eqn:infl_objective_grad_4}
    \begin{aligned}
      \del{ \Psi^{\rm Infl} (\inflvec) }{\inflfac_i} 
      &=  \inflfac_i^{-1} \,  \vec{e}_i \tran  \widetilde{\mat{A}}   \vec{e}_i
      - \inflfac_i^{-1} \,  \vec{e}_i \tran  \mat{H}\tran \left( \mat{R} + \mat{H} \widetilde{\mat{B}} \mat{H}\tran  \right)^{-1}  \mat{H}  \widetilde{\mat{B}} \widetilde{\mat{A}}   \vec{e}_i \,.
    \end{aligned}
  \end{equation}
  The derivative~\eqref{eqn:infl_objective_grad_4} can be efficiently evaluated given the matrices of ensemble anomalies
  of both the forecast and analysis, respectively.
  Moreover, we can proceed further as follows to avoid constructing the analysis ensemble at every iteration:
  \begin{equation}\label{eqn:infl_objective_grad_5}
    \widetilde{\mat{A}} 
    = \left( \mat{I} + \widetilde{\mat{B}} \mat{H}\tran \mat{R}^{-1} \mat{H} \right)^{-1} \widetilde{\mat{B}}
    = \widetilde{\mat{B}} - \widetilde{\mat{B}} \mat{H}\tran \left(\mat{R} + \mat{H} \widetilde{\mat{B}} \mat{H}\tran  \right)^{-1} \mat{H} \widetilde{\mat{B}}  \,.
  \end{equation}

  Then,
  \begin{equation}\label{eqn:infl_objective_grad_6}
    \begin{aligned}
      \widetilde{\mat{B}}^{-1}  \widetilde{\mat{A}}  \widetilde{\mat{A}}
      &= {\left( \mat{I} - \mat{H}\tran \left( \mat{R} + \mat{H} \widetilde{\mat{B}} \mat{H}\tran  \right)^{-1}  \mat{H}  \widetilde{\mat{B}} \right)}
      {\left( \widetilde{\mat{B}}  -  \widetilde{\mat{B}} \mat{H}\tran \left(\mat{R} + \mat{H} \widetilde{\mat{B}} \mat{H}\tran  \right)^{-1} \mat{H} \widetilde{\mat{B}} \right)} \\
      %
      %
      &= \widetilde{\mat{B}}  -  \widetilde{\mat{B}} \mat{H}\tran \mat{G}^{-1} \mat{H} \widetilde{\mat{B}} - \mat{H}\tran \mat{G}^{-1}  \mat{H}  \widetilde{\mat{B}}  \widetilde{\mat{B}}
    + \mat{H}\tran \mat{G}^{-1}  \mat{H}  \widetilde{\mat{B}}  \widetilde{\mat{B}} \mat{H}\tran \mat{G}^{-1} \mat{H} \widetilde{\mat{B}}  \,,
    \end{aligned}
  \end{equation}
  where $\mat{G} = \mat{R} + \mat{H} \widetilde{\mat{B}} \mat{H}\tran $.
  Now, by inserting~\eqref{eqn:infl_objective_grad_6} in~\eqref{eqn:infl_objective_grad_3}, the derivative reduces to
  \begin{equation}\label{eqn:A_Opt_Grad}
    \begin{aligned}
      \del{ \Psi^{\rm Infl} (\inflvec) }{\inflfac_i} 
      &=  \inflfac_i^{-1} \,  \vec{e}_i \tran \left(
      \widetilde{\mat{B}}
      -  \widetilde{\mat{B}} \mat{H}\tran \mat{G}^{-1} \mat{H} \widetilde{\mat{B}}
      - \mat{H}\tran \mat{G}^{-1}  \mat{H}  \widetilde{\mat{B}}  \widetilde{\mat{B}}
      + \mat{H}\tran \mat{G}^{-1}  \mat{H}  \widetilde{\mat{B}}  \widetilde{\mat{B}} \mat{H}\tran \mat{G}^{-1} \mat{H} \widetilde{\mat{B}}
      \right) \vec{e}_i \,.  
      %
      %
    \end{aligned}
  \end{equation}

  By defining $z_1 =  \widetilde{\mat{B}}  \vec{e}_i$,  $z_2 = \mat{H}\tran \mat{G}^{-1}  \mat{H}  \widetilde{\mat{B}}  z1$, 
  $z_3 = \widetilde{\mat{B}} \mat{H}\tran \mat{G}^{-1} \mat{H} z_1$, and $z_4 = \mat{H}\tran \mat{G}^{-1}  \mat{H}  \widetilde{\mat{B}} z_3$,
  we can  write the gradient of~\eqref{eqn:Infl_A_Opt_Objective} in the form
  %
  \begin{equation}\label{eqn:A_Opt_Grad_abbrev}
    \nabla_{\inflvec}{\Psi^{\rm Infl} (\inflvec)} 
    = \sum_{i=1}^{\Nstate}{ \vec{e}_i \, \del{ \Psi^{\rm Infl} (\inflvec) }{\inflfac_i}  }
    = \sum_{i=1}^{\Nstate}{ \inflfac_i^{-1}\, \vec{e}_i \vec{e}_i \tran \left( z_1  - z2 - z3 + z4 \right)} \,.
  \end{equation}

\section{OED Adaptive Localization}
\label{append:A_localization_gradient}
  Here, we derive the gradient of the A-optimality criterion for adaptive localization.

  \subsection{State-space gradient}
    The derivative of the objective~\eqref{eqn:B_localization_objective} is obtained as follows.
    For each $i=1,2,\ldots,\Nstate$
    \begin{equation} \label{eqn:loc_objective_grad_1}
        \del{ \Psi^{B-Loc}(\locvec)}{\locrad_i}
        %
        = \Trace{ - \widehat{\mat{A}} \del{ \widehat{\mat{A}}^{-1} } {\locrad_i}  \widehat{\mat{A}} } 
        = - \Trace{ \widehat{\mat{A}} \del{\left(   \widehat{\mat{B}}^{-1} + \mat{H}\tran \mat{R}^{-1} \mat{H} \right)}{\locrad_i} \widehat{\mat{A}} } 
        = - \Trace{ \widehat{\mat{A}} \del{ \widehat{\mat{B}}^{-1} }{\locrad_i} \widehat{\mat{A}} } \,,
    \end{equation}
    where $\widehat{\mat{A}}$ is a function of the vector of the localization radii $\locvec$.
    We proceed by finding the derivative of $\widehat{\mat{B}}^{-1}$, as follows:
    \begin{equation}\label{eqn:wtilde_B_deriv}
      \del{ \widehat{\mat{B}}^{-1} }{\locrad_i}
        = - \widehat{\mat{B}}^{-1}  \del{ \widehat{\mat{B}} }{\locrad_i} \widehat{\mat{B}}^{-1}
        = - \widehat{\mat{B}}^{-1}  \del{ \left( \mat{B} \odot \decorrmat \right) }{\locrad_i} \widehat{\mat{B}}^{-1} 
        = - \widehat{\mat{B}}^{-1}  \left( \mat{B} \odot \del{ \decorrmat }{\locrad_i} \right) \widehat{\mat{B}}^{-1}  \,.
    \end{equation}

This requires  finding the derivative of the decorrelation matrix,  $\del{ \decorrmat }{l_i} $.
    For that, let us define the following:
    \begin{equation}\label{eqn:loc_stride_vec}
         \mat{l}_i 
          = \left( \del{\decorrcoeff_{i,1}(\locrad_i)}{\locrad_i},
          \del{\decorrcoeff_{i,2}(\locrad_i)}{\locrad_i},
          \ldots,  
          \del{\decorrcoeff_{i,\Nstate}(\locrad_i)}{\locrad_i} 
          \right) \tran \,, 
    \end{equation}
    where the derivatives of the localization functions in~\eqref{eqn:loc_stride_vec} are defined as follows\footnotemark; for Gauss localization
    \begin{equation}\label{eqn:loc_fun_derivative_Gauss}
        \del{\decorrcoeff_{i,j}(\locrad_i)}{\locrad_i} = \frac{d(i, j)^2}{\locrad_i^3} \exp{\left( \frac{-d(i, j)^2}{2 \locrad_i^2} \right)} \,, 
    \end{equation}
    while for GC localization,
    \begin{equation}\label{eqn:loc_fun_derivative_GC}
        \del{\decorrcoeff_{i,j}(\locrad_i)}{\locrad_i} 
          = 
          \begin{cases}
            \frac{5}{4}  \frac{d(i,j)^5}{ \locrad_i^6 } - 2 \frac{d(i,j)^4}{\locrad_i^5} - \frac{15}{8} \frac{d(i,j)^3}{\locrad_i^4} + \frac{10}{3} \frac{d(i,j)^2}{\locrad_i^3}
              \,, \quad  & 0 \leq d(i,j) \leq \locrad_i \\
            -\frac{5}{12} \frac{d(i,j)^5}{\locrad_i^6} + 2 \frac{d(i,j)^4}{\locrad_i^5}  -  \frac{15}{8} \frac{d(i,j)^3}{\locrad_i^4}
              - \frac{10}{3} \frac{d(i,j)^2}{\locrad_i^3} + 5 \frac{d(i,j)}{\locrad_i^2}
              -\frac{2}{3} \frac{1}{d(i,j)} \,,  \quad & \locrad_i \leq d(i,j) \leq 2\locrad_i \\
            0\,. \quad & 2\locrad_i \leq d(i,j)
          \end{cases}
        \,.
    \end{equation}
    \footnotetext{Here, we assume $i\neq j$, since the distance metric satisfies $d(i,\,j)=0$ for $i=j$ and consequently $ \del{\decorrcoeff_{i,i}}{\locrad_i} = 0 $.}

    The following two identities will be useful in the following derivations.
    \begin{equation} \label{eqn:loc_objective_grad_3}
      \begin{aligned}
      \widehat{\mat{B}}^{-1} \widehat{\mat{A}} &= \widehat{\mat{B}}^{-1} \left( \widehat{\mat{B}}^{-1} + \mat{H}\tran \mat{R}^{-1} \mat{H} \right)^{-1}
        = \left( \mat{I} + \mat{H}\tran \mat{R}^{-1} \mat{H} \widehat{\mat{B}} \right)^{-1}  \,, \\
      \widehat{\mat{A}} \widehat{\mat{B}}^{-1} &= \left( \widehat{\mat{B}}^{-1} + \mat{H}\tran \mat{R}^{-1} \mat{H} \right)^{-1} \widehat{\mat{B}}^{-1}
        = \left( \mat{I} + \widehat{\mat{B}}  \mat{H}\tran \mat{R}^{-1} \mat{H} \right)^{-1}  \,.
      \end{aligned}
    \end{equation}

    Now, we can proceed with $\mat{B} \odot \del{ \decorrmat }{l_i}$, based on the specific choice of the decorrelation
    matrix $\decorrmat$ as defined by~\eqref{eqn:decorrelation_matrix}.
    Assuming $\decorrmat = \frac{1}{2} \left( \decorrmat_r + \decorrmat_r \tran \right)$,
    and using~\eqref{eqn:loc_stride_vec}, we have
    \begin{equation} \label{eqn:loc_objective_grad_4}
      \begin{aligned} 
      \mat{B} \odot \del{ \decorrmat_r }{l_i}
        &= \mat{B} \odot  \left(\vec{e}_i \mat{l}_i\tran \right)
        =  \vec{e}_i  \left( \mat{l}_i\tran \odot \left(\vec{e}_i \tran \mat{B} \right) \right) 
        = \vec{e}_i \,\mat{l}_{B,i}  \,, 
        \\
      \mat{B} \odot \del{ \decorrmat_r \tran }{l_i}
        &= 
          \left( \mat{B} \odot \del{ \decorrmat_r }{l_i} \right) \tran = \mat{l}_{B,i}\tran \, \vec{e}_i\tran \,,
      \end{aligned}
    \end{equation}
      where $ \mat{l}_{B,i} =  \mat{l}_i\tran \odot \left(\vec{e}_i \tran \mat{B} \right) $, and $\vec{e}_i$ is the $i$th cardinality vector in $\Rnum^{\Nstate}$.
      Then,
      \begin{equation}\label{eqn:B_dot_deriv}
        \mat{B} \odot \del{ \decorrmat }{l_i}
        = \mat{B} \odot \left( \frac{1}{2}( \vec{e}_i \mat{l}_i\tran +   \mat{l}_i \vec{e}_i\tran )  \right)
        = \frac{1}{2} \vec{e}_i  \mat{l}_{B,i} + \frac{1}{2} \mat{l}_{B,i}\tran \vec{e}_i\tran  \,.
      \end{equation}

      By combining results from~(\ref{eqn:loc_objective_grad_1},~\ref{eqn:wtilde_B_deriv},~\ref{eqn:loc_objective_grad_3},~\ref{eqn:B_dot_deriv}),
      and utilizing the circular property of matrix trace, we have
      \begin{equation}\label{eqn:loc_objective_grad_9}
        \begin{aligned}
         \del{ \Psi^{B-Loc}(\locvec)}{\locrad_i}
          &= \Trace{ \widehat{\mat{A}} \widehat{\mat{B}}^{-1} \left(\frac{1}{2} \vec{e}_i \mat{l}_{B,i} + \frac{1}{2} \mat{l}_{B,i}\tran \vec{e}_i\tran \right)  \widehat{\mat{B}}^{-1} \widehat{\mat{A}} }
          %
          = \Trace{ \mat{l}_{B,i} \widehat{\mat{B}}^{-1} \widehat{\mat{A}} \, \widehat{\mat{A}}  \widehat{\mat{B}}^{-1} \vec{e}_i} \\
          & = \mat{l}_{B,i} \widehat{\mat{B}}^{-1} \widehat{\mat{A}} \, \widehat{\mat{A}} \widehat{\mat{B}}^{-1} \vec{e}_i
          = \mat{l}_{B,i} {\left( \mat{I} + \mat{H}\tran \mat{R}^{-1} \mat{H} \widehat{\mat{B}} \right)^{-1}}
            {\left( \mat{I} + \widehat{\mat{B}}  \mat{H}\tran \mat{R}^{-1} \mat{H} \right)^{-1}} \vec{e}_i \,.
          %
          %
          %
        \end{aligned}
      \end{equation}
      %

  \subsection{Observation-space gradient}
  In this section, we derive the gradient of the A-optimality criterion for adaptive localization, formulated in the observation space.

    \subsubsection{Localizing $\mathbf{H} \mat{B}$}
      \label{append:R_localization_gradient_1}
    Here we derive the gradient of the objective~\eqref{eqn:R_localization_A_opt_goal_1}. Recall the A-optimal objective:
    \begin{equation}\label{eqn:R_localization_A_opt_goal_1_2}
         \Psi^{R-Loc}(\locvec)
        = \Trace{\mat{B}} - \Trace{ \widehat{\mat{HB} }\tran { \left( \mat{R} + {\mat{HB} \mat{H}\tran} \right)}^{-1} \widehat{\mat{HB} } } \,.
    \end{equation}
    We note that even if the covariance localization is applied to the covariance  matrix $\mat{B}$,
    the diagonal of $\mat{B}$ does not change, and in this case $\Trace{\widehat{\mat{B}}} = \Trace{\mat{B}}$.
    Hence we can drop the first term since it is a constant w.r.t the localization radii $\mat{L}$.
    %
    By employing the product rule of derivatives, the derivative of~\eqref{eqn:R_localization_A_opt_goal_1_2} is obtained as follows.
    For each $ i=1,2,\ldots,Nobs$,
    \begin{equation}\label{eqn:loc_objective_grad_17}
      \begin{aligned}
        \del{\Psi^{R-Loc}(\locvec)}{\locrad_i}
        %
        &= - \Trace{
           \widehat{\mat{HB} }\tran  
             {\left( \mat{R} + {\mat{HB} \mat{H}\tran} \right)}^{-1} \del{\widehat{\mat{HB}}}{l_i}
            +  \del{  \widehat{\mat{HB}}\tran }{l_i} \, 
               { \left( \mat{R} + {\mat{HB} \mat{H}\tran} \right)}^{-1} \widehat{\mat{HB} } 
        } \\
        %
        %
        &= - 2 \Trace{ \widehat{\mat{HB} }\tran  
          {\left(\mat{R} + {\mat{H}\mat{B}\mat{H}\tran} \right)}^{-1} \, \del{ \widehat{\mat{HB} } }{l_i}  } \,.
      \end{aligned}
      \end{equation}

    Now, we define\footnote{Here the $\decorrcoeff_{i,j}(\locrad_i)$ refers to the localization coefficient calculated between $i$th observation grid point
    and the $j$th model grid point, using the localization radius associated with the $i$th observation grid point.}
    \begin{equation}\label{eqn:Is_slice_vec}
      \mat{l}^s_i 
        = \left( \del{\decorrcoeff_{i,1}(\locrad_i)}{\locrad_i},
                \del{\decorrcoeff_{i,2}(\locrad_i)}{\locrad_i},
                \ldots,  
                \del{\decorrcoeff_{i,\Nstate}(\locrad_i)}{\locrad_i} 
          \right) \tran 
        \,; i = 1,2, \ldots, \Nobs \,,
    \end{equation}
    %
    and we let $\mat{l}_{\rm HB,i} = \mat{l}^s_i \odot \left( \vec{e}_i\tran \mat{HB} \right)\tran $,
    with $\vec{e}_i$ being the $i$th cardinality vector in $\Rnum^{\Nobs}$.
    Then
    \begin{equation} \label{eqn:HB_derivative}
      \del{}{l_i} \left[ \widehat{\mat{HB} } \right]
      = \del{\decorrmat^{\rm loc, 1} }{l_i} \odot \mat{HB}
      = \vec{e}_i \left( {\mat{l}^s_i}\tran \odot \left( \vec{e}_i\tran \mat{HB} \right) \right)
      = \vec{e}_i \, \mat{l}_{\rm HB,i}\tran \,.
    \end{equation}
    %
    %

    By substituting~\eqref{eqn:HB_derivative} in~\eqref{eqn:loc_objective_grad_17}, and using the circular property of the trace,
    it follows that the derivative of~\eqref{eqn:R_localization_A_opt_goal_1_2} reduces to
    \begin{equation}\label{eqn:loc_objective_grad_18}
      \del{\Psi^{R-Loc}(\locvec)}{\locrad_i}
      %
      %
      = - 2 \, \mat{l}_{\rm HB,i}\tran \, \widehat{\mat{HB} }\tran { \left( \mat{R} + {\mat{HB} \mat{H}\tran} \right)}^{-1} \, \vec{e}_i \,.
      %
    \end{equation}
    %
    %
    %

    \subsubsection{Localizing $\mathbf{H} \mat{B}$
      and $\mathbf{HB}\mathbf{H}\tran$}
      \label{append:R_localization_gradient_2}
    The A-optimal objective~\eqref{eqn:R_localization_A_opt_goal_2} in this case is
    \begin{equation}\label{eqn:loc_objective_grad_19}
         \Psi^{R-Loc}(\locvec)
        = \Trace{\mat{B}} - \Trace{ \widehat{\mat{HB} }\tran {\left( \mat{R} + \reallywidehat{\mat{HB} \mat{H}\tran} \right)}^{-1} \widehat{\mat{HB} }} \,,
    \end{equation}
    %
    %
    %
    with a derivative of the following form.
    \begin{equation} \label{eqn:loc_objective_grad_19_deriv}
    \del{\Psi^{R-Loc}(\locvec)}{\locrad_i} =
    -\Trace{
        \del{}{\locrad_i}\left[ \widehat{\mat{HB} }\tran { \left( \mat{R} + \reallywidehat{\mat{HB} \mat{H}\tran} \right)}^{-1} \widehat{\mat{HB} } \right]
    }
    \end{equation}

    We proceed by using~\eqref{eqn:HB_derivative} and by utilizing the product rule of derivatives
    to the matrix derivative in the right-hand side of~\eqref{eqn:loc_objective_grad_19_deriv}, as follows:
    \begin{equation}\label{eqn:loc_objective_grad_21}
      \begin{aligned}
        \Trace{
        \del{}{\locrad_i}\left[ \widehat{\mat{HB} }\tran { \left( \mat{R} + \reallywidehat{\mat{HB} \mat{H}\tran} \right)}^{-1} \widehat{\mat{HB} } \right]
        }
      %
        %
        %
        &= \Trace{\widehat{\mat{HB} }\tran \del{}{l_i}  { \left( \mat{R} + \reallywidehat{\mat{HB} \mat{H}\tran} \right)}^{-1} \, \widehat{\mat{HB} } }
        + 2 \,\mat{l}_{\rm HB,i}\tran\,   \psi^o_i  \,,
      \end{aligned}
    \end{equation}
    where
    \begin{equation}
      \psi^o_i = \widehat{\mat{HB} }\tran  { \left( \mat{R} + \reallywidehat{\mat{HB} \mat{H}\tran} \right)}^{-1}  \,  \vec{e}_i \,.
    \end{equation}
Formulating the derivative of~\eqref{eqn:loc_objective_grad_19} w.r.t $\locrad_i$
    requires the derivative
    \begin{equation}\label{eqn:loc_objective_grad_22}
      \begin{aligned}
        \del{}{l_i} { \left( \mat{R} + \reallywidehat{\mat{HB} \mat{H}\tran} \right)}^{-1}
        &= - { \left( \mat{R} + \reallywidehat{\mat{HB} \mat{H}\tran} \right)}^{-1}
        \del{{ \left( \mat{R} + \reallywidehat{\mat{HB} \mat{H}\tran} \right)} }{l_i}
        { \left( \mat{R} + \reallywidehat{\mat{HB} \mat{H}\tran} \right)}^{-1}  \\
        %
        %
        &= - { \left( \mat{R} + \reallywidehat{\mat{HB} \mat{H}\tran} \right)}^{-1}
        \left( \left({\mat{HB} \mat{H}\tran}\right) \odot \del{ \decorrmat^{o|o} }{l_i} \right)
        { \left( \mat{R} + \reallywidehat{\mat{HB} \mat{H}\tran} \right)}^{-1} \,.
      \end{aligned}
    \end{equation}
    %

    The final step is to formulate the derivative of $\decorrmat^{o|o} $. To do so, we define
    \footnote{Here the $\decorrcoeff_{i,j}(\locrad_i)$ refers to the localization coefficient calculated between $i$ and $j$th observation grid points,
     using the localization radius associated with the $i$th observation grid point.}
    \begin{equation}\label{eqn:loc_stride_vec_obs}
         \mat{l}^o_i
          = \left( \del{\decorrcoeff_{i,1}(\locrad_i)}{\locrad_i},
                    \del{\decorrcoeff_{i,2}(\locrad_i)}{\locrad_i},
                    \ldots,
                    \del{\decorrcoeff_{i,\Nobs}(\locrad_i)}{\locrad_i}
             \right) \tran \,,  \\
    \end{equation}
   By assuming that $\vec{e}_i$ is the $i$th cardinality vector in $\Rnum^{\Nobs}$,
    the derivative of $\decorrmat^{o|o} $, based on its prespecified form~\eqref{eqn:decorrelation_matrix_obs}, refines to
    \begin{equation}\label{eqn:decorrelation_matrices_derivatives_obs}
      \del{ \decorrmat^{o|o} }{l_i} =
        \frac{1}{2} \left( \vec{e}_i \left(\mat{l}^o_i\right)\tran +   \mat{l}^o_i \vec{e}_i\tran \right) \,.
    \end{equation}

Substituting~\eqref{eqn:decorrelation_matrices_derivatives_obs} in~\eqref{eqn:loc_objective_grad_22} , we have
  \begin{equation}\label{eqn:loc_objective_grad_23}
      \del{}{l_i} \left[ { \left( \mat{R} + \reallywidehat{\mat{HB} \mat{H}\tran} \right)}^{-1} \right]
      = -  \frac{1}{2} { \left( \mat{R} + \reallywidehat{\mat{HB} \mat{H}\tran} \right)}^{-1}
      \, \left(\vec{e}_i \mat{l}^o_{B,i}  + \left( \vec{e}_i \mat{l}^o_{B,i} \right)\tran  \right) \,
      { \left( \mat{R} + \reallywidehat{\mat{HB} \mat{H}\tran} \right)}^{-1} \,,
  \end{equation}
  with
  \begin{equation}
    \mat{l}^o_{B,i} =  \left(\mat{l}^o_i\right)\tran \odot \left({\vec{e}_i\tran \mat{HB} \mat{H}\tran}\right) \,,
  \end{equation}
  where $\mat{l}^o_i$ is defined in~\eqref{eqn:loc_stride_vec_obs}, and we made use of the fact that
  $\left( \vec{e}_i \vec{z}\tran \right) \odot \mat{U} = \vec{e}_i \left(\vec{z}\tran \odot \left(\vec{e}_i\tran\mat{U}\right) \right) $ for a vector $\vec{z}$ and a matrix $\mat{U}$ of conformable shapes.
  Using~\eqref{eqn:loc_objective_grad_23} and the cyclic property of the trace, we have
      \begin{equation}\label{eqn:loc_objective_grad_23_2}
        \begin{aligned}
          \Trace{ \widehat{\mat{HB} }\tran \del{}{l_i} { \left( \mat{R} + \reallywidehat{\mat{HB} \mat{H}\tran} \right)}^{-1} \, \widehat{\mat{HB} } }
          %
          %
          &= - \Trace{ \widehat{\mat{HB} }\tran {\left( \mat{R} + \reallywidehat{\mat{HB} \mat{H}\tran} \right)}^{-1} \,
            \vec{e}_i \, \mat{l}^o_{B,i} \, { \left( \mat{R} + \reallywidehat{\mat{HB} \mat{H}\tran} \right)}^{-1}  \widehat{\mat{HB}} }  \\
          &= - \mat{l}^o_{B,i}\, { \left( \mat{R} + \reallywidehat{\mat{HB} \mat{H}\tran} \right)}^{-1}
            \widehat{\mat{HB} } \widehat{\mat{HB} }\tran { \left( \mat{R} + \reallywidehat{\mat{HB} \mat{H}\tran} \right)}^{-1} \, \vec{e}_i  \\
          &= - \eta^o_i \, \psi^o_i \,,
        \end{aligned}
      \end{equation}
      where
  \begin{equation}\label{eqn:loc_objective_grad_23_4}
  \begin{aligned}
    %
      %
    \eta^o_i &= \mat{l}^o_{B,i}\, { \left( \mat{R} + \reallywidehat{\mat{HB} \mat{H}\tran} \right)}^{-1}   \widehat{\mat{HB} } \,.
  \end{aligned}
  \end{equation}

  By combining~(\ref{eqn:loc_objective_grad_21},~\ref{eqn:loc_objective_grad_23_2}, and ~\ref{eqn:loc_objective_grad_23_4})
  along with~\eqref{eqn:loc_objective_grad_19_deriv},
  and using the cyclic property of the matrix trace,
  the derivative of~\eqref{eqn:loc_objective_grad_19} follows: 
  \begin{equation}\label{eqn:loc_objective_grad_20_2}
    \begin{aligned}
      \del{\Psi^{R-Loc}(\locvec)}{\locrad_i}
      %
          &= \left( \eta^o_i - 2\, \mat{l}_{\rm HB,i}\tran  \right)\,  \psi^o_i \,.
    \end{aligned}
  \end{equation}
  %


\section*{Acknowledgments}
This material is based upon work supported by the U.S.\ Department of
Energy, Office of Science, Advanced Scientific Computing Research
Program under contract DE-AC02-06CH11357. 

%
\bibliographystyle{plain}
\bibliography{Bib/data_assim_fdvar,Bib/data_assim_HMC,Bib/data_assim_general,Bib/data_assim_Kalman,Bib/DATeS_Software,Bib/comprehensive_bibliography,Bib/oed_references}
%
\null
\vfill

\begin{flushleft} \scriptsize
  \framebox{\parbox{0.35\textwidth}{
    This document is provided by the contributing author(s) as a means to ensure timely dissemination of scholarly and technical work on a noncommercial basis. Copyright and all rights therein are maintained by the author(s) or by other copyright owners. It is understood that all persons copying this information will adhere to the terms and constraints invoked by each author's copyright. This work may not be reposted without explicit permission of the copyright owner.
  }} \hfill
\framebox{\parbox{0.58\textwidth}{
The submitted manuscript has been created by UChicago Argonne, LLC, 
Operator of Argonne National Laboratory (``Argonne"). Argonne, a
U.S. Department of Energy Office of Science laboratory, is operated
under Contract No. DE-AC02-06CH11357. The U.S. Government retains for
itself, and others acting on its behalf, a paid-up nonexclusive,
irrevocable worldwide license in said article to reproduce, prepare
derivative works, distribute copies to the public, and perform
publicly and display publicly, by or on behalf of the Government.
The Department of
Energy will provide public access to these results of federally sponsored research in accordance
with the DOE Public Access Plan. http://energy.gov/downloads/doe-public-access-plan. }}
\normalsize
\end{flushleft}

\end{document}